\documentclass[aps,prx,twocolumn,psfig,showpacs,superscriptaddress]{revtex4}
\usepackage{times}
\usepackage{graphicx}
\usepackage{float}
\usepackage{latexsym,amsmath,amssymb,bm,euscript}
\usepackage{color}
\usepackage{subfigure}
\usepackage{epstopdf}
\usepackage[colorlinks=true,linkcolor=blue,citecolor=blue]{hyperref}
\usepackage{hyperref}
\usepackage{soul}
\usepackage{ulem}
\usepackage{mathrsfs}
\usepackage{amsmath}

\sethlcolor{yellow}
\begin{document}

\title{Criticality-Enhanced Magnetocaloric Effect in Quantum Spin Chain Material Copper Nitrate}

\author{Junsen Xiang}
\affiliation{Department of Physics, Key Laboratory of Micro-Nano Measurement-Manipulation and Physics (Ministry of Education), Beihang University, Beijing 100191, China}

\author{Cong Chen}
\affiliation{Department of Physics, Key Laboratory of Micro-Nano Measurement-Manipulation and Physics (Ministry of Education), Beihang University, Beijing 100191, China}

\author{Wei Li}
\email{w.li@buaa.edu.cn}
\affiliation{Department of Physics, Key Laboratory of Micro-Nano Measurement-Manipulation and Physics (Ministry of Education), Beihang University, Beijing 100191, China}
\affiliation{International Research Institute of Multidisciplinary Science, Beihang University, Beijing 100191, China}

\author{Xianlei Sheng}
\affiliation{Department of Physics, Key Laboratory of Micro-Nano Measurement-Manipulation and Physics (Ministry of Education), Beihang University, Beijing 100191, China}
\affiliation{Department of Physics and Astronomy, University of Delaware, Newark, Delaware 19716-2570, USA}

\author{Na Su}
\affiliation{State Key Laboratory of Magnetism and Beijing National Laboratory for Condensed Matter Physics, Institute of Physics, Chinese Academy of Sciences, Beijing 100190, China}

\author{Zhaohua Cheng}
\affiliation{State Key Laboratory of Magnetism and Beijing National Laboratory for Condensed Matter Physics, Institute of Physics, Chinese Academy of Sciences, Beijing 100190, China}

\author{Qiang Chen}
\affiliation{Department of Physics, Key Laboratory of Micro-Nano Measurement-Manipulation and Physics (Ministry of Education), Beihang University, Beijing 100191, China}

\author{Ziyu Chen}
\email{chenzy@buaa.edu.cn}
\affiliation{Department of Physics, Key Laboratory of Micro-Nano Measurement-Manipulation and Physics (Ministry of Education), Beihang University, Beijing 100191, China}

\begin{abstract}
Low-dimensional quantum magnets, due to the existence of abundant exotic quantum phases therein and experimental feasibilities in laboratories, continues intriguing people in condensed matter physics. In this work, a comprehensive study of Cu(NO$_3$)$_2$ $\cdot$ 2.5H$_2$O (copper nitrate hemipentahydrate, CN), a spin chain material, is performed with multi-technique approach including thermal tensor network (TTN) simulations, first-principles calculations, as well as magnetization measurements in experiments. Employing a cutting-edge TTN method developed in the present work, we determine the couplings $J=5.13$ K, $\alpha=0.23(1)$ and Land\'e factors $g_{\parallel}=2.31$, $g_{\perp}=2.14$ in an alternating Heisenberg antiferromagnetic chain model, with which one can fit strikingly well the magnetothermodynamic properties. Part of the fitted experimental data are measured on the single-crystal CN specimens synthesized by us. Based on first-principles calculations, we reveal explicitly the spin chain scenario in CN by displaying the calculated electron density distributions, from which the distinct superexchange paths are visualized. On top of that, we investigated the magnetocaloric effect (MCE) in CN by calculating its isentropes and magnetic Gr\"ueisen parameter (GP). Prominent quantum-criticality-enhanced MCE was uncovered, the TTN simulations are in good agreements with measured isentropic lines in the sub-Kelvin region. We propose that CN is potentially a very promising quantum critical coolant, due to the remarkably enhanced MCE near both critical fields of moderate strengths as 2.87 and 4.08 T, respectively.
\end{abstract}

% 75.10.Jm     Quantized Spin Model
% 75.40.Cx     Static  properties  (order  parameter,  static  susceptibility,  heat
% 05.30.-d     Quantum Statistical Mechanics
% 05.30.Rt     Quantum Phase Transition
% 75.30.Sg     Magnetocaloric Effects
\maketitle

\section{Introduction}

Heisenberg spin chains and nets, owing to their strong quantum fluctuations and correlation effects, can accommodate plentiful interesting quantum phases like topological spin liquids \cite{Balents-2010, Yan-2011}, unconventional excitations like anyon-type quasi particles \cite{Rahmani-2014}, and inspiring behaviors like Bose-Einstain condensation in magnets \cite{Zapf-2014}, which continues stimulating both condensed matter theorists and experimentalists. What is more, these low-dimensional systems, which at a first glance are of purely academic interest, can actually have their experimental realizations. People have successfully discovered and/or synthesized plenty of spin materials which are very well described by the low-dimensional Heisenberg-type spin models. The long list includes, to name only a few, the diamond spin chain material azurite \cite{Kikuchi-2005}, the kagome spin liquid herbertsmithite \cite{Han-2012}, and copper nitrate as an alternating Heisenberg antiferromagnetic chain (AHAFC) \cite{Berger-1963, Friedberg-1968, Garaj-1968, Amaya-1969, Myers-1969, Morosin-1970, Van-1971, Van-1972, Van-1973, Amaya-1977, Diederix-1977, Diederix-1978a, Diederix-1978b, Diederix-1979, Eckert-1979, Bonner-1983, Taylor-1986, Xu-1999, Tennant-2003, Morozov-2003, Stone-2014, Willenberg-2015}. Therefore, low-dimensional quantum magnets arouses long-lasting research interest both theoretically and experimentally.

Among many other interesting properties of low-dimensional quantum magnets, we emphasize the enhanced magnetocaloric effect (MCE) in quantum critical regime. MCE is an intrinsic property of magnetic materials which exploits the reversible entropy changes caused by varying magnetic fields. MCE has a long history of study \cite{Warburg-1881, Weiss-1917, Smith-2013}, and in the past decades, developing novel MCE materials which have prominent MCE properties, like the Gadolinium alloys with giant MCE \cite{Gschneidner-1997, Smith-2012}, has raised great research interest. This is due to that MCE has appealing applications in eco-friendly refrigeration near room temperature \cite{Brown-1976, Pecharsky-1997}, providing a good substitute to conventional vapor compression refrigeration, and also in space technology \cite{Hagmann-1995, Zimm-2003}. In addition, MCE materials, in particular adiabatic dimagnetization refrigerant (ADR), serve as efficient coolants for ultra low temperature (sub-Kelvin) regime \cite{Giauque-1927, Giauque-1933, Rost-2009, Sharples-2014}. People pursues MCE refrigerant which have higher isothermal entropy change ($\Delta S$), larger adiabatic temperature difference ($T_{\rm{ad}}$), and also lower hysteresis dissipation \cite{Smith-2012}.

Recently, quantum spin chain materials are shown to exhibit enhanced MCE even at ultra low temperatures, and thus raised great research interest \cite{Ryll-2014, Zhu-2003, Garst-2005, Rost-2009, Sharples-2014, Lucia, Zhitomirsky-2004, Honecker-2009, Lang-2010, Wolf-2011}. On one hand, through exploring low-$T$ MCE properties of spin chain model materials \cite{Ryll-2014, Rost-2009} which shows divergent Gr\"uneisen parameter near field-induced quantum critical points (QCP), people are able to directly detect and study quantum criticality \cite{Zhu-2003, Garst-2005}. On the other hand, one can inversely utilize this low-temperature thermodynamic anomaly to realize enhanced cooling effects near quantum critical points (QCPs) \cite{Zhitomirsky-2004, Honecker-2009}. Very recently, Sharples \textit{et al.} realized temperatures as low as $\sim$ 200 mK using the enhanced MCE of a molecular quantum magnet \cite{Sharples-2014}, and Lang \textit{et al.} experimentally studied a spin-1/2 Heisenberg antiferromagnetic chain material [Cu($\mu$-C$_2$O$_4$)(4-aminopyridine)$_2$(H$_2$O)]$_n$ (CuP, for short) \cite{Lang-2010}, and demonstrated this quantum critical coolant is a perfect alternative to standard ADR salts, due to its wider operating temperature range, longer hold time and high efficiency.\cite{Wolf-2011}

In order to study the thermodynamic information including the appealing MCE property of these strongly correlated spin systems, accurate thermal algorithms are of crucial significance, which is indispensable in establishing links between theoretical spin models and experimental measurements at finite temperatures. In one spatial dimension (1D), the transfer matrix renormalization group (TMRG) method \cite{TMRG} has been long accepted as the method of reference, owing to its high accuracy and versatility. In Ref. \onlinecite{LTRG}, Li \textit{et al.} proposed an alternative approach for calculating thermodynamics of low-dimensional quantum lattice models called linearized tensor renormalization group (LTRG) method, which also adopts the Trotter-Suzuki decomposition \cite{Suzuki-1976} to express the partition function as a thermal tensor network (TTN) and linearly contract the resulting $d+1$ dimensional ($d=1,2$ for 1D and 2D lattice, respectively) TTN along Trotter direction via renormalization group (RG) techniques.

In this work, combining three different methods, i.e., thermal quantum manybody computation, first-principles calculations, and experimental measurements of magnetization, we performed a comprehensive investigation of an alternating quantum spin chain material Cu(NO$_3$)$_2$ $\cdot$ 2.5H$_2$O (copper nitrate hemipentahydrate, hereinafter referred to as ``CN"). CN is one of the earliest spin chain material ever studied experimentally \cite{Berger-1963,Friedberg-1968,Garaj-1968,Amaya-1969,Myers-1969,Morosin-1970, Van-1971, Van-1972, Van-1973, Amaya-1977, Diederix-1977, Diederix-1978a, Diederix-1978b, Diederix-1979, Eckert-1979, Bonner-1983, Taylor-1986, Xu-1999, Tennant-2003, Morozov-2003, Stone-2014, Willenberg-2015, Wittekoek-1968}, while continues intriguing people for its abundant physics including triplon wave excitation \cite{Xu-1999} and precise Tomanaga-Lutting liquid behavior \cite{Willenberg-2015}. We notice that, despite many efforts, discrepancy in coupling constants still exists: the exact diagonalization (ED) fittings ($J=5.16$ K, $\alpha=0.27$) to thermodynamic measurements measurably deviates from those obtained from inelastic neutron scattering (INS) experiments ($J=5.14$ K, $\alpha=0.24$) \cite{Xu-1999}.

In order to resolve this discrepancy in couplings, we generalize the LTRG approach to a bilayer formulation (dubbed as LTRG++) which further improves the accuracy of calculations. With this cutting-edge TTN method at hand, we revisit the previous experimental data in Ref. \onlinecite{Bonner-1983} including specific heat curves (at various fields) and magnetization curves, augmented with magnetization measurements done in this work. The couplings are determined precisely to be $J=5.13$ K, $\alpha=0.23(1)$, $g_{\parallel}=2.31$, and $g_{\perp}=2.14$, consistent with that from INS experiments. In addition, first-principles calculations present electron density distributions and therefore visualized superexchange paths, thus providing direct and indubitable proof on the spin-chain alignment in material CN. Furthermore, through TTN simulations, we show that CN has large entropy change and pronounced peaks (and dips) in Gr\"uneisen parameter around QCPs at low temperatures, and the calculated adiabatic temperature changes can fit strikingly well to previously measured isentropes, revealing that CN may be a remarkably ideal quantum critical refrigerant.

The rest of the article is arranged as followings: Section \ref{Sec:CNMaterial} presents a brief overview of the previous experimental and theoretical studies on material CN. An \textit{ab initio} study is also carried out which reveal intra- and inter-chain couplings directly and explicitly; Section \ref{Sec:ThTNAlg} is devoted to a review of the TTN algorithms and the development of bilayer LTRG; In Section \ref{Sec:ThTNFit} we show the TTN simulations and its fitting to experiments data, including both taken from previous measurements and done in the present work; Section \ref{Sec:MCE} discusses the criticality-enhanced MCE in material CN and its potential usage as low-temperature coolant; Lastly, Section \ref{Sec:Summary} presents the conclusion and discusses some possible extensions of the present work.

\begin{figure*}[tbp]
  \centering
  \includegraphics[angle=0,width=0.9\linewidth]{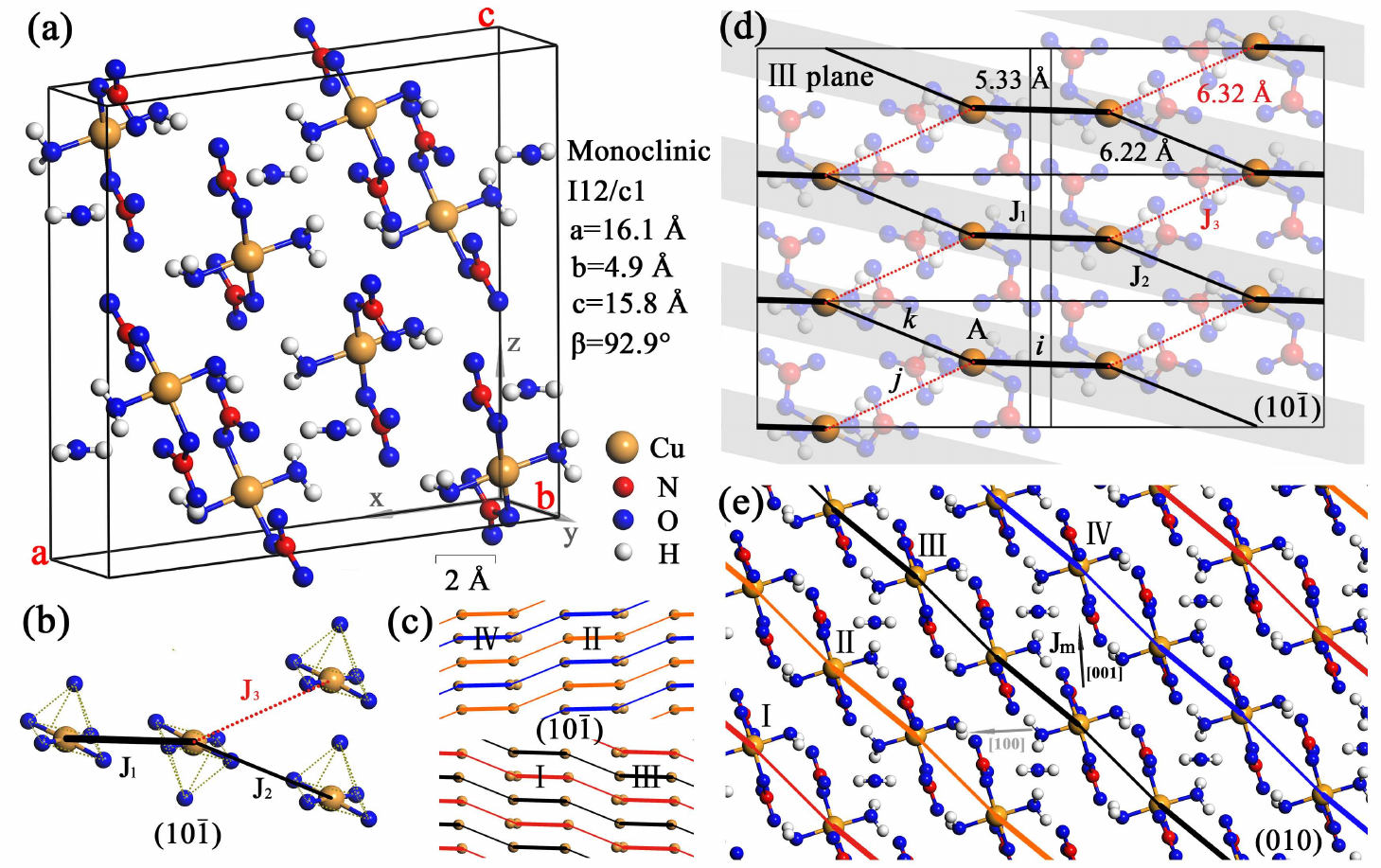}
  \caption{(Color online) Crystal structure and magnetic exchange couplings
  in Cu(NO$_3$)$_2$ $\cdot$ 2.5H$_2$O. (a) The unit cell of Cu(NO$_3$)$_2$
  $\cdot$ 2.5H$_2$O, where the coordinate axes coincide with the crystal
  axes. The lattice constants are shown in the figure, indicating that CN belongs to the
  monoclinic system. (b) The expanded view of the quare pyramidal
  polyhedrons comprise of oxygen atoms and Cu$^{2+}$ ions. (c) Superexchange paths between spins along chains in four inequivalent ($10\bar{1}$) planes which are adjacent to each other. (d) Structure in a typical (10$\bar{1}$) plane, where the Cu$^{2+}$ ions are highlighted while other atoms left transparent. The heavy solid lines are the intradimer interactions $J_1$, the interdimer $J_2$ and interchain $J_3$ couplings are plotted differently (in black solid and red dashed lines, respectively). $A$ labels one out of two sublattices of honeycomb structure in (10$\bar{1}$) plane, and $\hat{i}, \hat{j}, \hat{k}$ are vectors connecting one site (in $A$ sublattice) with its three nearest neighbors. (e) Projected view of the crystal structure in (010) plane, where the alternating solid lines represents the $J_1$-$J_2$ chain. We denote the four existing (10$\bar{1}$) planes as
  \uppercase\expandafter{\romannumeral1}\,
  \uppercase\expandafter{\romannumeral2}\,
  \uppercase\expandafter{\romannumeral3}\  and
  \uppercase\expandafter{\romannumeral4}, respectively, where the chains have different paths in each plane. Arrows indicate the
  directions along [100] and [001], which represent interchain exchange path $J_L$, $J_m$. }
  \label{Fig:CN}
\end{figure*}

\section{Alternating Heisenberg Antiferromagnetic Spin Chain Material Copper Nitrate}
\label{Sec:CNMaterial}

Copper nitrate hemipentahydrate [Cu(NO$_3$)$_2$ $\cdot$ 2.5H$_2$O] was the first inorganic S=1/2 spin chain material studied experimentally \cite{Berger-1963,Friedberg-1968}. As one of the common copper salts, CN possesses some special thermodynamic properties at the low temperatures, including the zero-magnetization plateau \cite{Myers-1969, Diederix-1979}, 1D Luttinger liquid behavior under magnetic fields \cite{Sakai-1995,Willenberg-2015}, and 3D magnetic transition at ultra low temperature (150 $\sim$ 160 mK) \cite{Amaya-1977, Diederix-1978b,Willenberg-2015}, etc, which has aroused people's research interest for more than half a century, significantly promoting developments of the research on low-dimensional quantum magnets.

Before diving into detailed discussions on CN, we point that there exist other copper nitrate hydrates. Among others, the Cu(NO$_3$)$_2$ $\cdot$ 3H$_2$O (copper nitrate trihydrate) has similar formulas and exactly the same X-ray diffraction pattern \cite{Schrein-1924}, and thus is easy to be confused with CN hemipentahydrate studied here. We paid special attention to discriminate between these two similar hydrates, and according to thermogravimetric analysis \cite{Taylor-1986,Morozov-2003}, it is concluded that Cu(NO$_3$)$_2$ $\cdot$ 2.5H$_2$O is a distinct hydrate different from Cu(NO$_3$)$_2$ $\cdot$ 3H$_2$O.

\subsection{Crystal Structure}
Figure \ref{Fig:CN} depicts the crystallographic structure of CN, which is monoclinic with space group $I12/c1$ \cite{Garaj-1968}. The corresponding lattice constants are found to be $a$ = 16.1 \AA, $b$ = 4.9 \AA, $c$ =15.8 \AA, and $\beta$ = 92.9 $^\circ$ at low temperatures ($\sim$ 3 K) \cite{Xu-1999, Tennant-2003}. As shown in Fig. \ref{Fig:CN}(a), a conventional unit cell comprises of eight formula units. Each Cu$^{2+}$ ion is surrounded by five nearest-neighboring oxygen atoms, which constitutes a distorted pyramidal polyhedron [Fig \ref{Fig:CN}(b)]. Four oxygen atoms resides at the vertices of the basal plaquette of the polyhedron, roughly on the same plane: two of these oxygen atoms belong to H$_2$O molecules and the other two are from NO$^{-}_{3}$ groups; the rest apical oxygen atom belongs to a third NO$^{-}_{3}$ group. The pyramidal polyhedrons of opposite orientations are arranged alternatively along a line [Fig. \ref{Fig:CN}(b)].

\subsection{Alternating Heisenberg Antiferromagnetic Chain}
\label{SubSec:AHAFC}
Based on early experimental research on CN, including the measurements of magnetic susceptibility \cite{Berger-1963} and specific heat \cite{Friedberg-1968}, a binary cluster model for describing its magnetic properties was proposed: Cu$^{2+}$ ions having spin $S=1/2$ are coupled in pair (with coupling strength $\frac{1}{2} J/k_B=2.56$ K) \cite{Berger-1963}, and the system thus comprises of independent spin binary clusters. In addition, people also perform proton magnetic resonance (PMR) experiments and confirm that the short-range antiferromagnetic order are within spin-pairs, thus supporting the dimer model. \cite{Wittekoek-1968}

Although the binary cluster model can capture some of the main features of spin-spin correlation in CN, discrepancy between this simple model and experimental measurements still remains. A weak interdimer exchange interaction ($J_2$) was then brought into the binary cluster model in Ref. \onlinecite{Myers-1969}, which substantially improves the fitting to the isothermal magnetization curve. In addition, van Tol, et al., studied the magnetic phase transitions via PMR and uncovered a crossover to short-range magnetic order at 350 mK, as well as a critical transition to long-range order at 160 mK, under a magnetic field of 3.6 T, which can be understood only by introducing inter-dimer couplings in the model \cite{Van-1971}. Besides the PMR data, this is also evidenced in the specific heat measurement, where the low temperature hump(peak) signals the crossover(transition) into short-(long-) range ordered state \cite{Van-1973,Diederix-1978a}, also suggesting the existence of weak inter-dimer couplings.

Early attempts to introduce inter-dimer interactions include two possible model structures: a ladder model (with dimers on the rungs) and an alternating chain model \cite{Tachiki-1970}. These two models both fit thermodynamic measurements of CN well since their corresponding thermal predictions are essentially equal \cite{Diederix-1978a, Bonner-1983}. However, the discrimination between these two models was later done by the angular dependent PMR \cite{Diederix-1978b}, according to which one identifies the possible superexchange paths and thus rules out the ladder model. In addition, neutron-diffraction results also support the spin chain scenario \cite{Eckert-1979}.

Therefore, it has been concluded that an AHAFC model can very well describes the magnetic properties of CN (in the temperature regime above $\sim$ 160 mK), which reads
\begin{equation}
H = J \sum_{n = 1}^{L/2} (\vec{S}_{2n-1} \vec{S}_{2n} + \alpha \vec{S}_{2n} \vec{S}_{2n+1}) - \sum_{m=1}^{L} \sum_{\nu=\{\parallel, \perp\}} g_{\nu} B_{\nu} S_m^z,
\label{Eq-Hamiltonian}
\end{equation}
where $\vec{S} = \{ S^x, S^y, S^z \}$ is the vector spin operator containing three spin operators in different directions; $J=J_1$ is the strongest superexchange coupling; $\alpha=J_2/J_1$ is the relative strength of dominant inter-dimer interaction, whose precise value was measurably different in various experiments and left undetermined between $0.24$ and $0.27$ \cite{Bonner-1983, Xu-1999}. Also note that in the magnetic-field coupling (Zeeman) term, the Land\'e factors are different ($g_{\parallel} \neq g_{\perp}$) on the direction along b axis and that perpendicular to it. This magnetic anisotropy has been observed experimentally in the magnetic susceptibility measurements for a period of time \cite{Berger-1963}, while little understanding has been achieved yet.

Figures \ref{Fig:CN}(b-e) depict the spin chain structure and the spin-spin interaction paths. The distances between one Cu$^{2+}$ ion to three neighbors are 5.33 \AA\ , 6.22 \AA\ , and  6.32 \AA\ \cite{Diederix-1978a}, which leads to, through Cu-O-H-O-N-O-Cu bridges, three distinct couplings $J_1$, $J_2$, and $J_3$, respectively [Fig. \ref{Fig:CN}(b)]. Therefore two possible inter-dimer superexchange paths $J_2$ and $J_3$ are shown in Fig. \ref{Fig:CN}(d), and the spin chain thus could have had two possible routes on any $(10\bar{1})$ plane [see Fig\ref{Fig:CN}(d)]. Until recently, inelastic neutron scattering (INS) determines that $J_3$=-0.01 meV (of magnitude about 1/10 of $J_2$) \cite{Stone-2014}, so that $J_2$ is the dominant inter-dimer interaction, which connects dimers to form a tilted alternating chain, as shown n Figs. \ref{Fig:CN}(c-e).

Moreover, from Fig. \ref{Fig:CN}(e), we can see that there exist four inequivalent types of $(10\bar{1})$ planes where the spin chains are arranged in different ways, namely, planes I to IV shown in Fig. \ref{Fig:CN}(e). In I and III planes, the AHAFC stretch along $[111]$ direction [from left top to right bottom, see Fig. \ref{Fig:CN}(c)], with a shift of $b/2\simeq$ 2.45 \AA\ between chains in I and III planes; while in planes II and IV, the chains go from left bottom to right top ($[1\bar{1}1]$ direction), with the same shift of distance ($b/2$) between spin chains in both planes [Fig. \ref{Fig:CN}(c)].

\subsection{First-Principles Calculation and Electron Density Distributions}
\label{SubSec:Abintio}
In Subsection \ref{SubSec:AHAFC}, we scrape together quite a number of experimental observations, including various thermodynamic measurements and INS results, and arrive at the conclusion of an AHAFC model description for CN. However, a thorough study of electronic structures in CN via \textit{ab initio} calculations is indispensable, which may provide an direct check for the existence of spin-chain type magnetic interactions in CN and offer insight into exchange path other than intra-chain couplings.

In this work, we employ a self-consistent field calculation, based on the all-electron projector augmented wave (PAW) method~\cite{paw} implemented in VASP \cite{Kresse-1993}, to investigate the electron density distributions in CN. We adopt the generalized gradient approximation of Perdew, Burke, and Ernzerhof exchange-correlation functional \cite{Perdew-1996}. The cutoff energy for the plane wave expansion is chosen as 1000 eV, and the k-point mesh is 2$\times$3$\times$2. In practical calculations, little changes both in the cell shape and atomic positions have been obseved after structure relaxation, hence the experimental lattice parameters shown in Fig. 1(a) are used, and two unit cells which comprise 264 atoms (including 16 copper atoms) are selected.

\begin{figure}[tbp]
  \centering
  \includegraphics[angle=0,width=0.9\linewidth]{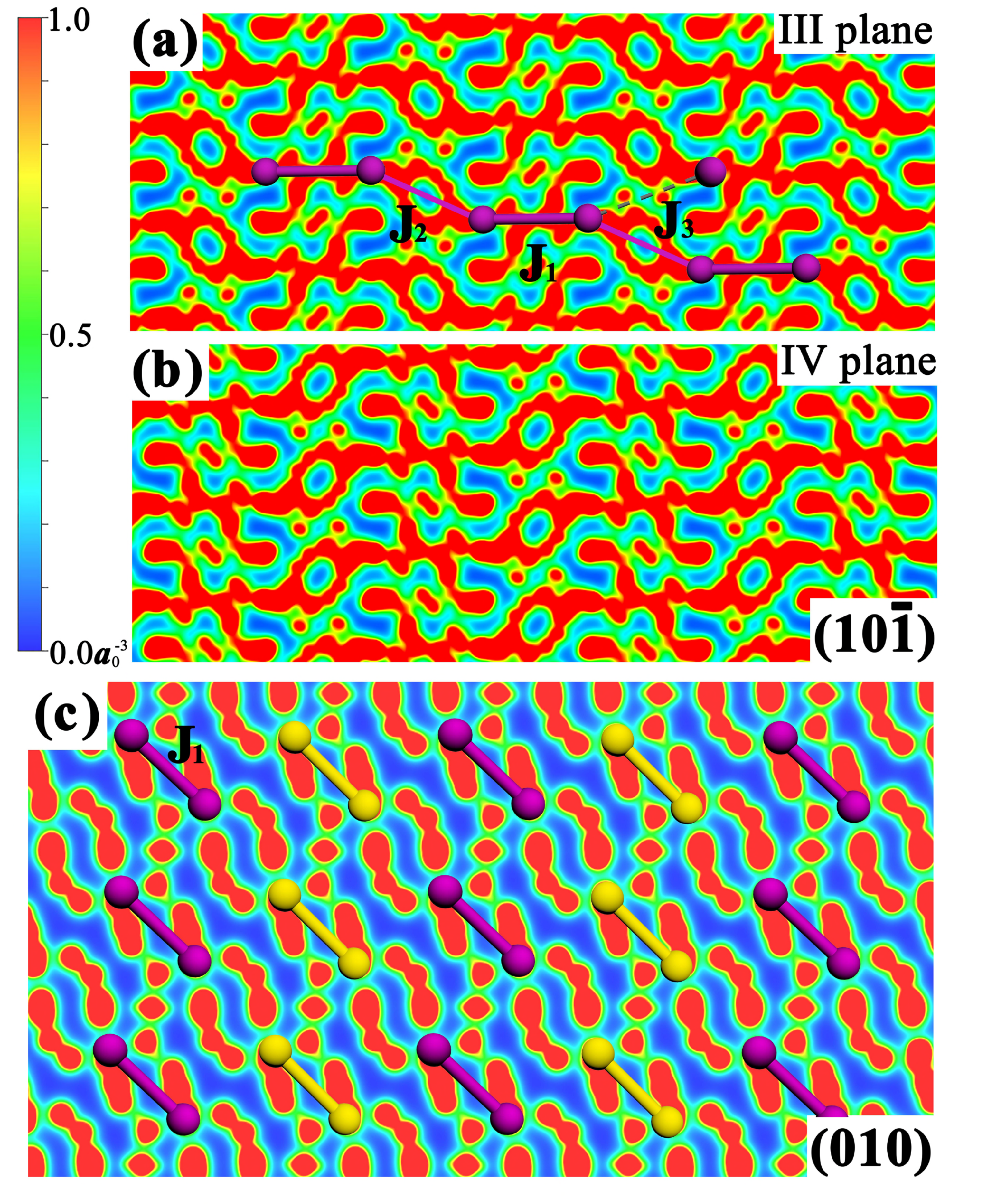}
  \caption{(Color online) The projected electron densities on (a) III-type (10$\bar{1}$), (b) IV-type (10$\bar{1}$), and (c) (010) planes. $a_0 \simeq 0.53$ \AA\, is the Bohr radius, the projection range of electron density is of thickness [-0.5, 0.5], respect to [10$\bar{1}$] unit vector for (a,b) and to [010] vector (i.e., primitive vector $b$) for (c) (refer to Fig. \ref{Fig:CN} for the specific crystal directions). The positions of copper ions are marked by solid balls. In (a,b) the tilted chain structures are clearly shown by high electron densities along the chain direction [111] for (a) and [1$\bar{1}1$] for (b). In (c) the dimers with different hight on $b$ axis are labeled in different colors, from which it is clear that there exist weak inter-dimer interactions (denoted as $J_m$) along [001] direction, while there exists no visible exchange path between two nearest neighboring dimers connected by 1/2 $\times$ [111] or [1$\bar{1}$1] vector.}
  \label{Fig:EDD}
\end{figure}

In Fig. \ref{Fig:EDD}, we show the simulated results of electron density distributions. Remarkably, in Figs. \ref{Fig:EDD}(a,b) the spin chain alignment in (10$\bar{1}$) plane is clearly demonstrated, where the electrons tend to reside along the chain directions and thus leads to larger exchange integrals $J_1$ and $J_2$ [see Fig. \ref{Fig:CN}(a)]. Note that from the calculated results, we can discriminate $J_2$ from $J_3$ without any ambiguity, where the Fig. \ref{Fig:EDD} show that the electron densities (hence also the coupling strength) are different in $J_2$ and $J_3$ bonds for orders of magnitudes. This difference between $J_2$ and $J_3$, as well as the fact that tilted chains are along difference directions between I, III and II, IV planes [Figs. \ref{Fig:EDD}(a,b)], agree with the INS observations in Refs. \onlinecite{Xu-1999,Stone-2014}. Moreover, in Fig. \ref{Fig:EDD}(c) we show the electron densities in (010) plane, where the $J_1$ dimers are highlighted, from which we can see that there exists a weak dimer-dimer exchange coupling $J_m$ between every pair of dimers along [001] direction, this again has been observed experimentally \cite{Xu-1999}.

\subsection{Inter-chain Couplings and 3D Heisenberg spin model}
The AHAFC model can explain well most experimental observations of CN in a very wide range of temperatures (above a few hundreds of milikelvins). Nevertheless, ultra low temperature measurements of PMR \cite{Van-1971}, adiabatic susceptibility \cite{Diederix-1977}, and heat capacity \cite{Bonner-1983}, uncovered a 3D phase transition at about $150\sim160$ mK, under magnetic fields between $B_{c}$ and $B_{s}$. The existence of such a finite-temperature transition from magnetic disordered phase to 3D long-range order phase \cite{Willenberg-2015} is clearly beyond the 1D spin model given in Eq. (\ref{Eq-Hamiltonian}).

In Subsection \ref{SubSec:AHAFC}, we mentioned that the ferromagnetic coupling $J_3$ whose magnitude is about 1/10 of $J_2$ [Figs. \ref{Fig:CN}(b,d)], and thus plays the role of a weak inter-chain interaction. In addition, INS experiments reveal that there exist some other inter-chain interactions, namely, $J_m = 0.018(2)$ meV, between dimers and along [001] directions [Fig \ref{Fig:CN}(e)] \cite{Xu-1999}. These important facts have been confirmed by our first-principles calculations in Subsection \ref{SubSec:Abintio}, as Fig. \ref{Fig:EDD} illustrates.

Put all these intra- and inter-chain interactions together, we arrive at a 3D Heisenberg spin model with Hamiltonian $H_{3D} = H_{(10\bar{1})} + H_{\rm{CTC}}$, which reads:
\begin{equation}
H_{(10\bar{1})} = \sum_{\vec{p} \in \{A\}} J_1 \vec{S}_{\vec{p}} \vec{S}_{\vec{p} + \hat{i}} + J_j \vec{S}_{\vec{p}} \vec{S}_{\vec{p} + \hat{j}} + J_k \vec{S}_{\vec{p}} \vec{S}_{\vec{p} + \hat{k}},
\label{Eq:HalmitonianHC}
\end{equation}
and
\begin{equation}
H_{\rm{CTC}} = \sum_{\vec{p}} J_{m} \vec{S}_{\vec{p}} \vec{S}_{\vec{p}+\vec{c}/2}
\label{Eq:HalmitonianCTC}
\end{equation}

Equation (\ref{Eq:HalmitonianHC}) is a 2D honeycomb lattice model with three different coupling constants along $\hat{i}$, $\hat{j}$, and $\hat{k}$ directions [see Fig. \ref{Fig:CN}(d)]. $\vec{p}$ is the coordination vector in $\{ A \}$ sublattice [Eq. \ref{Eq:HalmitonianHC}] or running over all lattice sites [Eq. \ref{Eq:HalmitonianCTC}] of honeycomb lattice on (10$\bar{1}$) planes. One typical site in $\{ A \}$ sublattice has been marked in Fig. \ref{Fig:CN}(d). Note that $J_j=J_2$ and $J_k=J_3$ on planes I and III, while $J_k=J_2$ and $J_j=J_3$ on planes II and IV.

Equation (\ref{Eq:HalmitonianCTC}) represent the inter-layer couplings, $J_m$ is the dimer coupling, $\vec{p}$ is coordination vector of every localized spin site, and $\vec{c}$ has magnitude of lattice constant $c=15.8$ \AA\ and is along [001] direction (there must be a spin site at $\vec{p} + \vec{c}/2$ according to CN structure shown in Fig. \ref{Fig:CN}). Some additional remarks on 3D inter-chain couplings in Eq. (\ref{Eq:HalmitonianCTC}) are in order: Since there exist four kinds of 2D honeycomb planes [I to IV in Figs. \ref{Fig:CN}(c,e)] which are different from each other by lattice shifts and reflection operations about (010) plane (see related descriptions in Section \ref{SubSec:AHAFC}), and the chains are arranged along different paths ([111] or $[11\bar{1}]$) on different planes, Eq. \ref{Eq:HalmitonianCTC} thus describes a 3D coupled tilted chains (CTC) model. In this model, two spins are coupled via $J_1$ (a dimer) when both involved sites are of equal height in b axis and have a shift in distance of $c/2 \simeq 7.9$ \AA\  in [001] direction in ac-plane. Therefore, the seemingly simple $H_{\rm{CTC}}$ in Eq. (\ref{Eq:HalmitonianCTC}) actually represents a quite peculiar 3D inter-chain coupling model which looks bizarre while are actually feasible in the material CN. This 3D model has not been reported before as far as we know, neither has its properties been explored.

In Ref. \onlinecite{Xu-1999}, INS experiments also show there exists inter-chain interaction between nearest dimers along [100] direction. However, we find that by shifting dimers along [1/2 0 0] as indicated by the authors in Ref. \onlinecite{Xu-1999}, there locates \textit{no} dimer in the supposed position (see Fig. \ref{Fig:CN}). This is also verified in our Abinitio calculations, where Fig. \ref{Fig:EDD}(c) shows clearly that there is no visible dimer-dimer coupling between a dimer and its nearest neighbor along [100] direction. Therefore, we include only the inter-dimer coupling $J_m$ along [001] direction in Eq. (\ref{Eq:HalmitonianCTC}), while leaving it as an open problem about the possibility of adding more inter-chain coupling terms to Eq. (\ref{Eq:HalmitonianCTC}). Note that the inter-chain interactions are rather weak and does not alter the physical properties except at ultra low temperatures. In the followings, except for the discussions on criticality-enhance MCE in CN, this 3D model Eqs. (\ref{Eq:HalmitonianHC}, \ref{Eq:HalmitonianCTC}) will not be involved, and we focus on the AHAFC model description in Eq. (\ref{Eq-Hamiltonian}) exclusively.

\begin{figure}[tbp]
  \includegraphics[angle=0,width=1\linewidth]{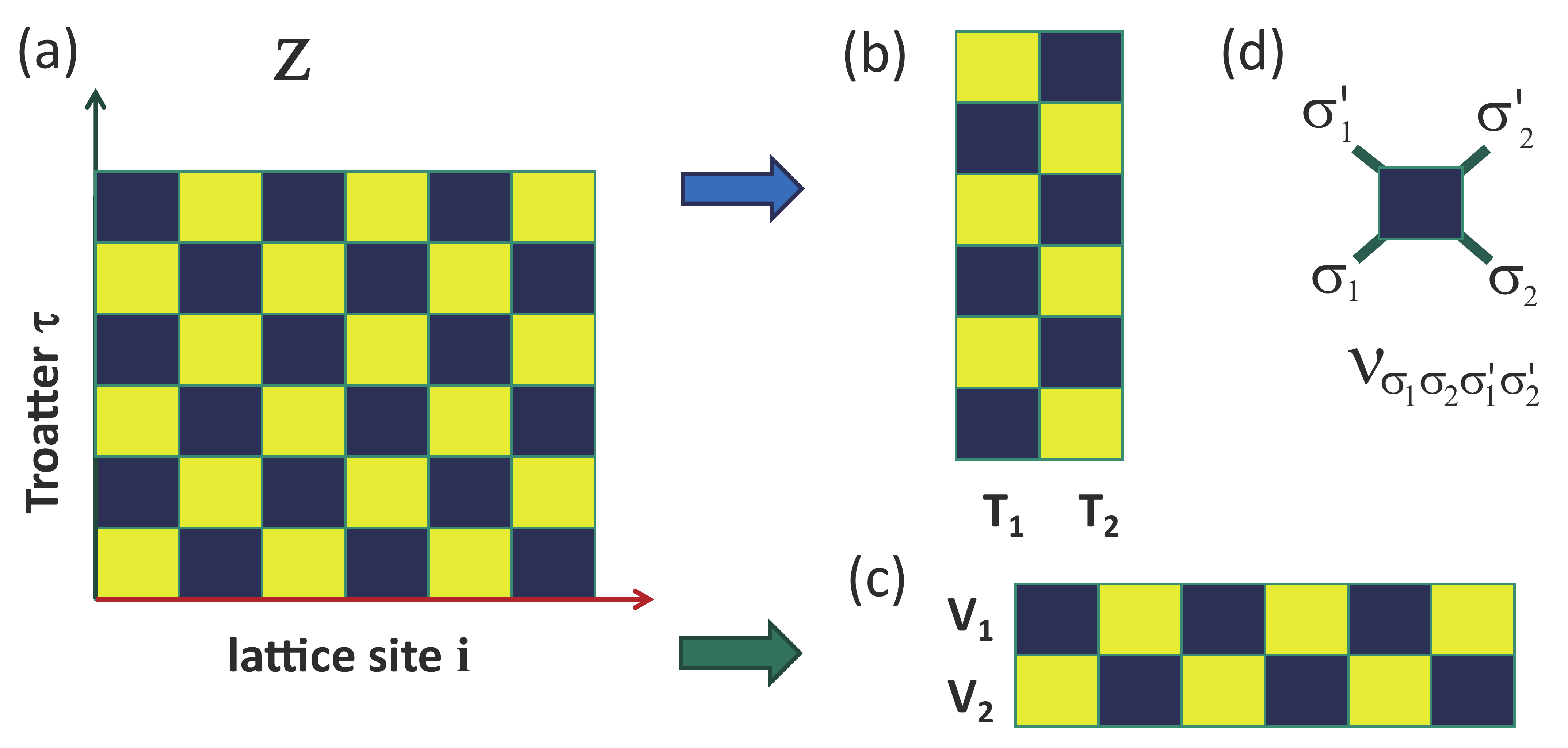}
  \caption{(Color online) (a) The 2D TTN represents the partition function $\bold{Z}$ of a 1D quantum lattice model, which exhibits a checkerboard pattern. (b) and (c) are the transfer matrices along spatial and Trotter directions, respectively. (d) depicts the rank-four local tensor $\nu$, the elementary unit in the TTN.}
  \label{Fig:TTN}
\end{figure}

\section{Thermal Tensor Network Approach}
\label{Sec:ThTNAlg}
High-precision thermal quantum manybody calculations are indispensable for relating the spin models discussed in Sec. \ref{Sec:CNMaterial} to the thermodynamical pmeasurements of CN. In this section, we firstly review the LTRG method proposed by some of the authors in Ref. \onlinecite{LTRG}. Then, we promote LTRG to a double-layer form with significantly improved accuracy, and then discuss its intrinsic relation to the well-established TMRG approach.

\subsection{Thermal Tensor Networks and Linearized Tensor Renormalization Group Method}
Employing the Trotter-Suzuki decomposition \cite{Suzuki-1976}, we obtain a 2D TTN for the 1D Heisenber chain, which consists of rank-four tensors $\nu_{\sigma_1, \sigma_2, \sigma_3, \sigma_4}=\langle \sigma_1, \sigma_2 | \exp{(-\tau h_{i,j})} | \sigma_3, \sigma_4 \rangle$ [see Figs. \ref{Fig:TTN} (d)]. In order to calculate the thermodynamic properties one needs to accurately contract the TTN, which, however, is a NP-hard problem and thus can not be solved exactly. Therefore, people has to resort to approximate methods for efficient contractions of TTN. Among others, renormalization group algorithms constitute an important class of approaches which are developed to accurately contract the TTN and calculate interested quantities including free energy per site, specific heat, magnetic susceptibility, entropy, and others.

As shown in Fig. \ref{Fig:TTN}, the interconnected local tensors constitute a 2D checkerboard-style TTN, which subsequentlyp can be regarded as repeated 1D vertical ($T_1,T_2$) or horizontal ($V_1, V_2$) stripes, i.e., transfer matrices, as shown in Figs. \ref{Fig:TTN} (b,c). The full contraction of TTN and consequently the calculations of thermal properties can be accomplished with the help of these transfer matrices. For instance, the vertical stripes $T_1,T_2$ transfers the spin indices $\{\sigma_i\}$ between different lattice sites and the partition function $Z$ thus reads
\begin{equation}
Z = \lim_{L \to \infty} \rm{Tr} (T_1 T_2)^{L/2},
\end{equation}
where $L$ denotes the total length of the chain. In the thermodynamic limit the dominant eigenvalue $\lambda_{\rm{max}}$ of transfer matrix $T=T_1 T_2$ determines the free energy per site $f = \frac{1}{2\beta} \ln \lambda_{\rm{max}}$ and also other thermodynamic quantities.

In order to calculate the extreme eigenvalue (and corresponding eigenvector), in Ref. \onlinecite{TMRG}, Xiang and Wang utilized the density-matrix renormalization group (DMRG) method, originally developed for Hamiltonian system, to solve the transfer matrix problem. They perform RG process along the Trotter direction and truncate the accumulated $\{\sigma_i\}$ indices/states into a fixed number ($M$) of renormalized states. TMRG can determine the thermodynamic properties with high precision, and has been established as the method of reference in calculating thermodynamics of 1D strongly correlated quantum lattice systems \cite{TMRGApp}.

Alternatively, the efficient contraction of the 1+1D TTN can also be performed by making use of the horizontal transfer matrices $V_1, V_2$ [see Fig. \ref{Fig:TTN} (c)]. Li \textit{et al.} proposed a TTN algorithm dubbed as LTRG \cite{LTRG}, which project continually the transfer matrix $V_{1(2)}$ to the density matrix of the system (in a form of matrix product operator, MPO). LTRG method can be used to accurately calculate the thermodynamics in 1D chains \cite{LTRG,Yan-2012} and applies also to higher dimensional lattices \cite{LTRG, Ran-2012}. At a first glance, it seems that these two RG methods, TMRG and LTRG, are quite different, since they are dealing with transfer matrices along two distinct axes, i.e., the horizontal and vertical directions, which are clearly inequivalent: the spatial direction is infinite while the thermometric axis is finite and subject to a periodic boundary condition. However, a closer look into it below reveals that they are actually intimately related to each other, on the ground that both methods manage to accurately contract the TTN, while such globally optimized contraction schemes should treat both directions (implicitly) in equal footing.

\subsection{LTRG and LTRG++ Algorithms}
\label{SubSec:LTRG++}
\begin{figure}[tbp]
  \includegraphics[angle=0,width=0.85\linewidth]{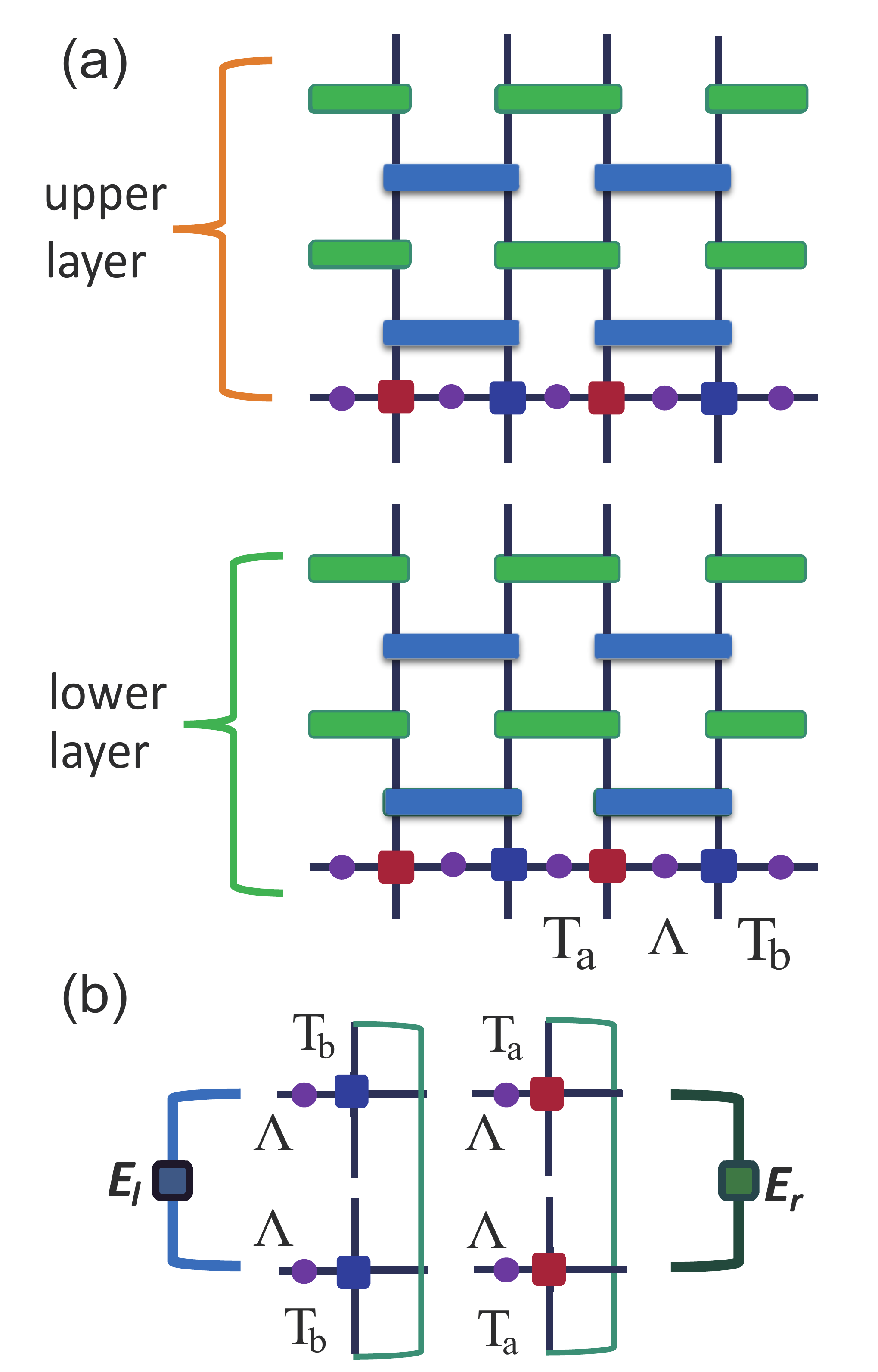}
  \caption{(Color online) LTRG++ algorithm adopts a double-layer scheme which contracts simultaneously upper and lower layers into two MPOs. (b) $T_{a(b)}$ and its conjugate counterpart constitute the transfer matrix, with the left(right) dominant eigenvector $E_{l(r)}$ and the corresponding eigenvalue $\lambda_{\rm{max}}$.}
  \label{Fig:LTRG++}
\end{figure}

In this subsection, we firstly briefly review the single-layer LTRG proposed in Ref. \onlinecite{LTRG}, and then propose its bilayer form (Fig. \ref{Fig:LTRG++}), which improves the accuracy prominently and is dubbed as LTRG++. For more technical details on LTRG, we refer to Appendix \ref{App:LTRG++}.

In the LTRG algorithm, only a single MPO [upper or lower one in Fig. \ref{Fig:LTRG++}(a)] is involved in the process of contractions: Starting from the single MPO at initial inverse temperature $\tau$, and by continually projecting $\nu$ tensors onto upper (or lower) MPO, we cool down the system from very high temperatures ($1/\tau$) to various lower temperatures, $\beta=1/((n+1)\tau)$ at $n$-th step and thus the thermodynamics at temperature $1/\beta$ can be calculated (see more details of single-layer LTRG in Ref. \onlinecite{LTRG}). LTRG can produce quite accurate results for thermodynamic properties even at quite low temperatures, and the precision is comparable to that of TMRG. \cite{LTRG}

Nevertheless, besides the single-layer algorithm, we devise here a double-layer LTRG++ algorithm, as shown in Fig. \ref{Fig:LTRG++}, for contracting the TTN. The main idea is as following: we contract the TTN into \textit{two} (instead of one) MPOs, this is what ``++" means. Figure \ref{Fig:LTRG++}(a) exploits a symmetric construction, where the upper and lower layers are contracted into two MPOs in exactly the same manner, saving one half of the projection time. Due to the checkerboard structure of TTN, each projection step comprises of two substeps, called as the odd and even substeps. After $n$ steps, one reaches the inverse temperature $\beta = (2n+2) \tau$. The free energy per site $f(\beta) = \lim_{N \to \infty} -\frac{1}{N} \log(Z)$ can now be calculated from the series of renormalization factors $\kappa^a$ and $\kappa^b$, extracted at odd and even substeps, respectively \cite{FootnoteKappa}. In addition, we also need to calculate the dominant eigenvalue $\lambda_{\rm{max}}$ of transfer matrix consisted of $T_a, T_b$ and their conjugates in Fig. \ref{Fig:LTRG++}(b), with corresponding left (right) eigenvectors $E_l$ ($E_r$). With these results, the free energy per site $f$ at $\beta = 2(n+1)\tau$ can be computed via
\begin{equation}
f = \frac{1}{4(n+1)} [\sum_{i=1}^{n} (\log \kappa_i^a + \log \kappa_i^b) + \log(\lambda_{\rm{max}})].
\label{Eq:LTRG++}
\end{equation}

The advantages of LTRG++ over the previous single-layer algorithm can be summarized in mainly two aspects: Firstly, LTRG++ further improves the accuracy of LTRG, and is now practically of the same precision as TMRG; Secondly, LTRG++ can save half of the projection time. Related discussions and numerical evidence can be found in Appendix \ref{App:LTRG++}. Note that in the LTRG++ algorithm one can also adopt an asymmetric construction such that two MPOs are (approximately) conjugate to each other, and recovers then the purification scheme \cite{Verstraete} used in finite-temperature DMRG \cite{Feiguin-2005}. Therefore, TTN approaches are very \textit{flexible}, and can be designed to rewrite/recover the well-established TMRG, purification, and potentially other thermal RG methods.

\section{Thermal tensor network simulation and fittings to CN experimental results}
\label{Sec:ThTNFit}

\begin{figure}[tbp]
  \includegraphics[angle=0,width=1.0\linewidth]{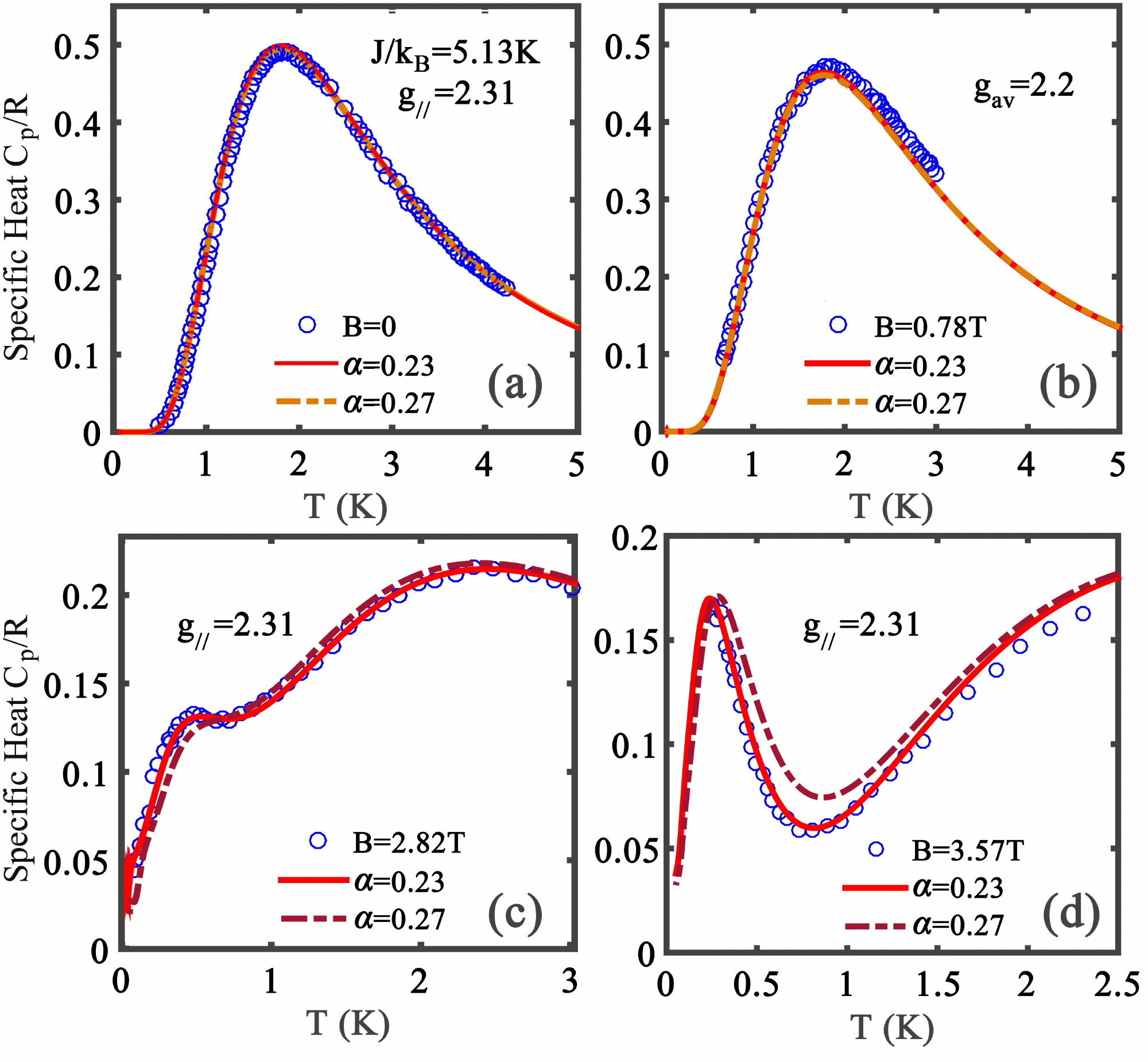}
  \caption{Fitting to experimental data of specific heat curves under various magnetic fields, (a) B = 0, (b) 0.78, (c) 2.82, and (d) 3.57 T. The experimental data (symbols) are taken from Refs. \onlinecite{Bonner-1983, Friedberg-1968}, and the dashed fitting lines are calculated with $\alpha=0.27$, while the solid lines are fittings with $\alpha=0.23$. The $B=0.78$ T curve in (b) was measured with powder samples \cite{Bonner-1983}, thus is fitted using average Land\'e factor $g_{\rm{av}} \simeq 2.2$; and the $B=2.82$ and 3.57 T curves in (c,d) are measured along crystal $b$ axis, with Land\'e factor $g_{\parallel} =2.31$.}
   \label{Fig:SpecificHeat} %% label for entire figure
\end{figure}

Fitting the numerical simulated results to experimentally obtained magnetothermodynamic data is an important way to determine the parameters in Hamiltonian Eq. (\ref{Eq-Hamiltonian}) describing the CN chain. In particular, in the present study, the purposes of performing numerical fittings are two-fold.

Firstly, we notice that fittings to the thermal properties includes the magnetic susceptibility, magnetization curves, and variousp specific heat curves have been performed in Ref. \onlinecite{Bonner-1983}, where part of the data can be nicely fitted based on ED calculations on system length up to $L=12$ (and their extrapolations). The authors get a set of fitting parameters, among which the coupling ratio between weak and strong bond is $\alpha=0.27$, and $J/k_B=5.16(4)$K) [see Hamiltonian Eq. (\ref{Eq-Hamiltonian})]. However, the specific heat curves at large magnetic fields (2.82 and 3.57 T) were poorly fitted by the ED calculations. Actually, since at such strong field (2.8 T $< B <$ 4.4 T) the ground state of the system is in a quantum critical region (Luttinger liquid phase \cite{Willenberg-2015}) and is supposed to have rather long correlation lengths at low temperatures. Therefore, ED calculations with such a small system size (up to $L=12$) might be insufficient for an accurate estimation in that circumstance, and a large-scale thermodynamic algorithm is essentially needed to clarify this ambiguity.

Secondly, and more importantly, it is also noticed that the coupling ration $\alpha = 0.27$ determined in Ref. \onlinecite{Bonner-1983} is ``measurably different" from that from INS experiments, where it is determined that $\alpha \simeq 0.24$ [with strong bond $J=0.442(2)$ meV = 5.13(2) K].\cite{Xu-1999} The authors in Ref. \onlinecite{Xu-1999} ascribed this discrepancy to the influence of interchain interactions in thermodynamic fittings. However, this argument can not be fully plausible since the interchain interaction is so weak ($\sim$ 0.06K) \cite{Diederix-1978a,Diederix-1978b} and could hardly induce any sizable effects on the thermodynamic properties in the temperature range of 2$\sim$20 K so as to shift the fitting from $\alpha=0.24$ to 0.27.

Therefore, we perform state-of-the-art TTN simulations developed in Sec. \ref{SubSec:LTRG++} and fit the magnetothermodynamic data both taken from Refs. \onlinecite{Berger-1963, Friedberg-1968, Bonner-1983} and measured in the present work. In our lab, we prepared various CN single-crystal specimens with sizes up to one or two centimeters (see Appendix .\ref{App:Sample} for details of sample preparation) and measured their magnetization in high-precision SQUID devices. In practical calculations, Trotter slice is set as $\tau=0.025$, the lowest temperature reached is $T/J=1/150$ (i.e., inverse temperature $\beta=150$), and $\chi=400\sim600$ bond states are retained, with truncation error smaller than $10^{-13}$. Numerical convergence versus $\chi$ of various concerned quantities including free energy, specific heat, magnetization curve, etc, has always been checked.

\subsection{Specific Heat at Various Magnetic Fields}
\label{SubSec:SpecificHeat}

We start from the specific heat curves at various magnetic fields $B$ = 0, 0.87, 2.82, and 3.57 T, as shown in Fig. \ref{Fig:SpecificHeat}. Experimental data (symbols) are taken from Ref. \onlinecite{Bonner-1983}, and the coupling constant $J$, chosen to be $5.13$ K, is within the fiducial range of 5.16(4) K from previous thermal fitting and 5.13(2) K from scattering fitting. Actually we find out that small change of $J$ (say, $\pm 0.3$ K) does not cause significant changes calculated results of quantities so as to affect other fitting parameters.

In Figs. \ref{Fig:SpecificHeat}(a), \ref{Fig:SpecificHeat}(b), we plot specific heat curves of low magnetic fields ($B$ = 0, 0.78 T), and both fittings with $\alpha=0.23$ (solid lines) and 0.27 (broken lines) are displayed as comparisons. It is seen clearly that the calculated curves of both $\alpha$ values can fit the magnetic specific heat curves almost equally well, for either $B=0$ or $B=0.78$ T case. Therefore, it is difficult making a preferable choice amongst these two $\alpha$ values, as well as potentially many other values in between.

In Fig. \ref{Fig:SpecificHeat}(c) the measured specific heat curve $C_p$ shows double peak structure, and $\alpha=0.24$ and 0.27 curves start to show some qualitatively different behaviors: While the $\alpha=0.27$ curve only presents a shoulder below 1 K, the calculated $\alpha=0.23$ curve correctly captures the double-peak structure, making the latter fitting noticeably better than the former. Moreover, the difference between two fittings becomes more strikingly in Fig. \ref{Fig:SpecificHeat}(d) \cite{Footnote1}, where $C_p$ in the region 0.3 $\sim$ 1.5 K is quite sensitive to the change of $\alpha$, and $\alpha=0.23$ is obviously superior than 0.27 in this case.

Therefore, from the direct comparisons in Fig. \ref{Fig:SpecificHeat}, we conclude that $\alpha=0.23$ is an overall better parameter than $\alpha=0.27$ in the fittings of specific heat curves at various fields. We would like to stress that the discrimination between $\alpha=0.23$ and 0.27 can be done only if an accurately calculation is possible for the low-temperature thermodynamic property of CN at high fields (2.82 and 3.57 T) where the ground state is a critical Luttinger liquid. The authors in Ref. \onlinecite{Bonner-1983} performed fittings to these data based ED simulations on rather limited system size (up to 12 sites) and get poor agreements, thus did not manage to discriminate $\alpha=0.24$ from $\alpha=0.27$ \cite{FootnoteCp}.

\subsection{Magnetic Susceptibility and Magnetization Curve}

\begin{figure}[tbp]
  \includegraphics[angle=0,width=1.0\linewidth]{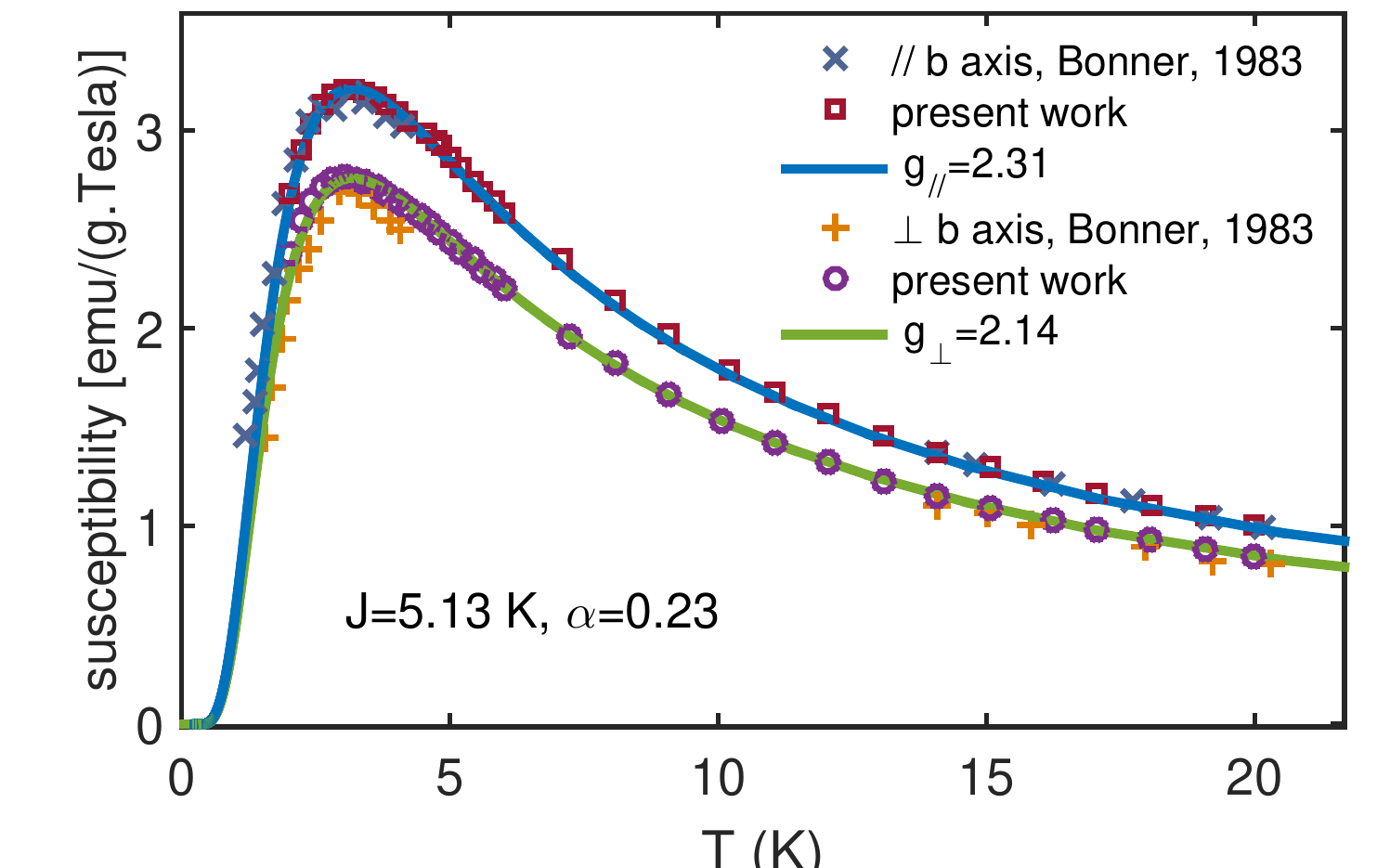}
  \caption{(Color online) Fittings to measured magnetic susceptibility $\chi$ taken from previous experiments (Ref. \onlinecite{Bonner-1983}), as well as those obtained in the present work (squares and circles). The latter is measured under a small magnetic field ($B=0.6$ T) to mimic the zero-field susceptibility. $\chi$ has clear annisotropic $g$ factors along the crystal $b$ axis and the direction perpendicular to it.}
  \label{Fig:Sus}
\end{figure}

\begin{figure}[tbp]
  \includegraphics[angle=0,width=1.0\linewidth]{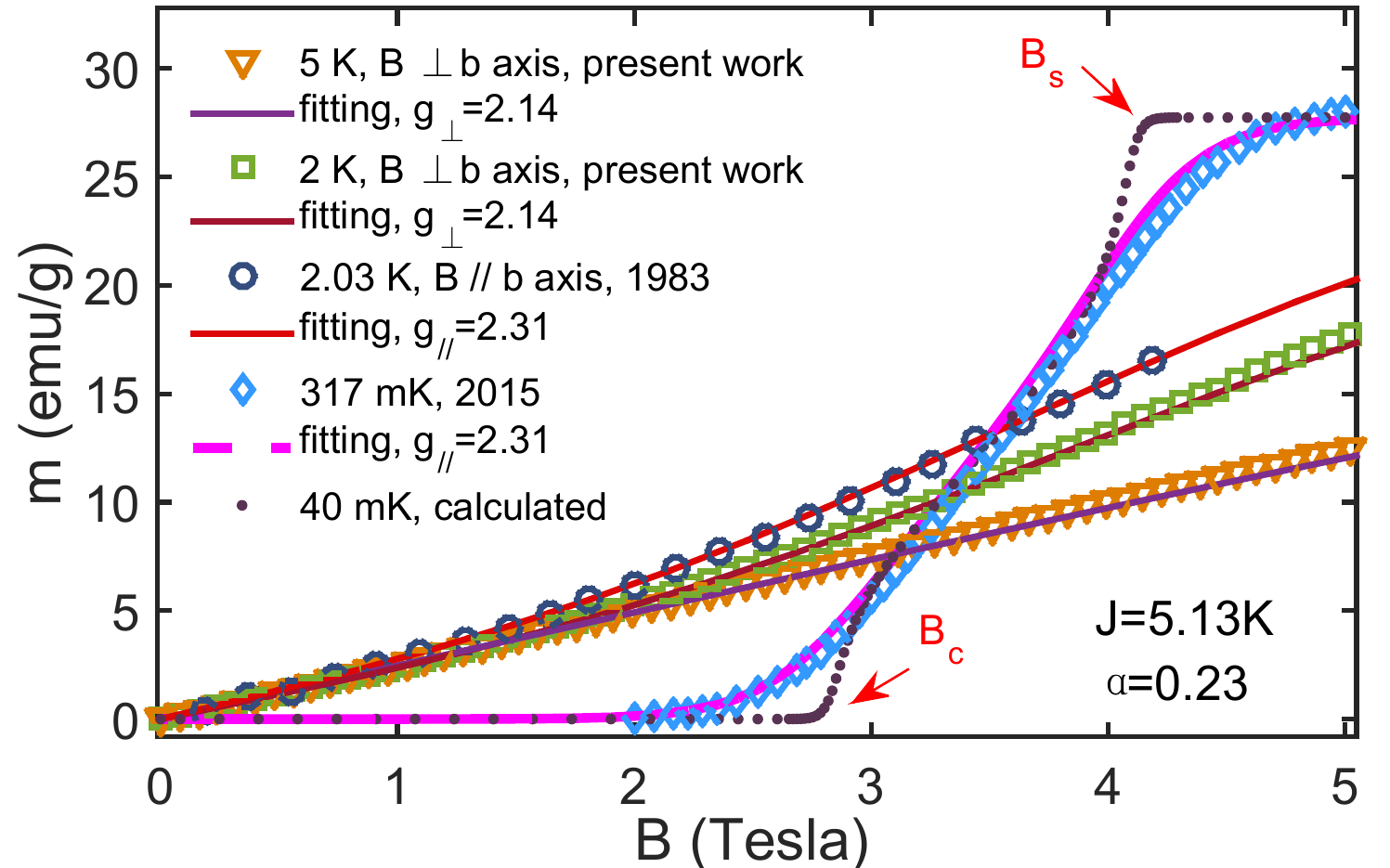}
  \caption{(Color online) Various magnetization curves at different temperatures and their LTRG fittings. The two curves (at 2 and 5 K) under magnetic fields applied perpendicular to the $b$ axis, are measured with a SQUID in the present work; while the 2.03 K curve parallel $b$ axis and 317 mK curve perpendicular to $b$ axis are taken from Ref. \onlinecite{Bonner-1983} and Ref. \onlinecite{Willenberg-2015}, respectively. A 40 mK line ideally calculated from the spin chain model is also included, demonstrating two quantum critical points $B_c=2.87$ T and $B_s=4.08$ T which are identified by two diverging peaks of $dM/dB$.}
  \label{Fig:MagCurves}
\end{figure}

In Sec. \ref{SubSec:SpecificHeat}, we find the parameter $\alpha=0.23$ prevails $\alpha=0.27$ for fitting specific heat curves under magnetic fields. In this subsection, we check whether the preferred parameter $\alpha=0.23$ can also fit other magnetothermodynamic quantities such as zero-field magnetic susceptibility measurements $\chi$ and the magnetization curves at various temperatures.

Figure \ref{Fig:Sus} illustrates the fittings to magnetic susceptibility reuslts, which comprises data measured in the present work and those taken from Ref. \onlinecite{Bonner-1983}. In particular, the present magnetic susceptibility measurements are performed in order to fill up the gap in the temperature range 5 K $< T <$ 15 K where the old susceptibility data are absent. It is seen that in Fig. \ref{Fig:Sus} the TTN calculations can fit the experimental results very well. Note that the magnetic susceptibilities are measured both along and perpendicular to the crystal $b$ axis, Fig. \ref{Fig:Sus} reveals that there exists quite prominent annisotropy in the spin chain material. It turns out, through the fittings of both susceptibilities with Hamiltonian Eq. \ref{Eq-Hamiltonian}, that this annisotropy can be attributed to different Laud\'e factors in the directions parallel ($g_{\parallel} = 2.31$) and perpendicular ($g_{\perp} = 2.14$) to the $b$ axis.

Besides the zero-field $\chi$, we also fitted the magnetization curves at various temperatures (517 mK, 2, 2.03, and 5 K). In Fig. \ref{Fig:MagCurves} the measured magnetization curves with magnetic fields perpendicular to b axis, parallel magnetization curves taken from Ref. \onlinecite{Bonner-1983}, as well as 517 mK data from Ref. \onlinecite{Willenberg-2015}, are quantitatively fitted with the set of parameters $J=5.13$ K, $\alpha=0.23$, $g_{\parallel}=2.31$, and $g_{\perp}=2.14$.

To summarize, through above high-precision fittings, we conclude that the coupling ratio $\alpha$ determined is close to that ($\alpha=0.24$) obtained from INS experiments \cite{Xu-1999}, while ``measurably" different from $\alpha=0.27$ in previous thermodynamic fittings \cite{Bonner-1983}. This finding reveals that the thermal and scattering experiments are actually \textit{consistent} with each other, and the previously supposed discrepancy may be due to limited simulations in fitting low-$T$ thermal data of gapless Luttinger liquid phase.

\section{Criticality-Enhanced Magnetocaloric Effect}
\label{Sec:MCE}

\begin{figure}[tbp]
  \includegraphics[angle=0,width=1.0\linewidth]{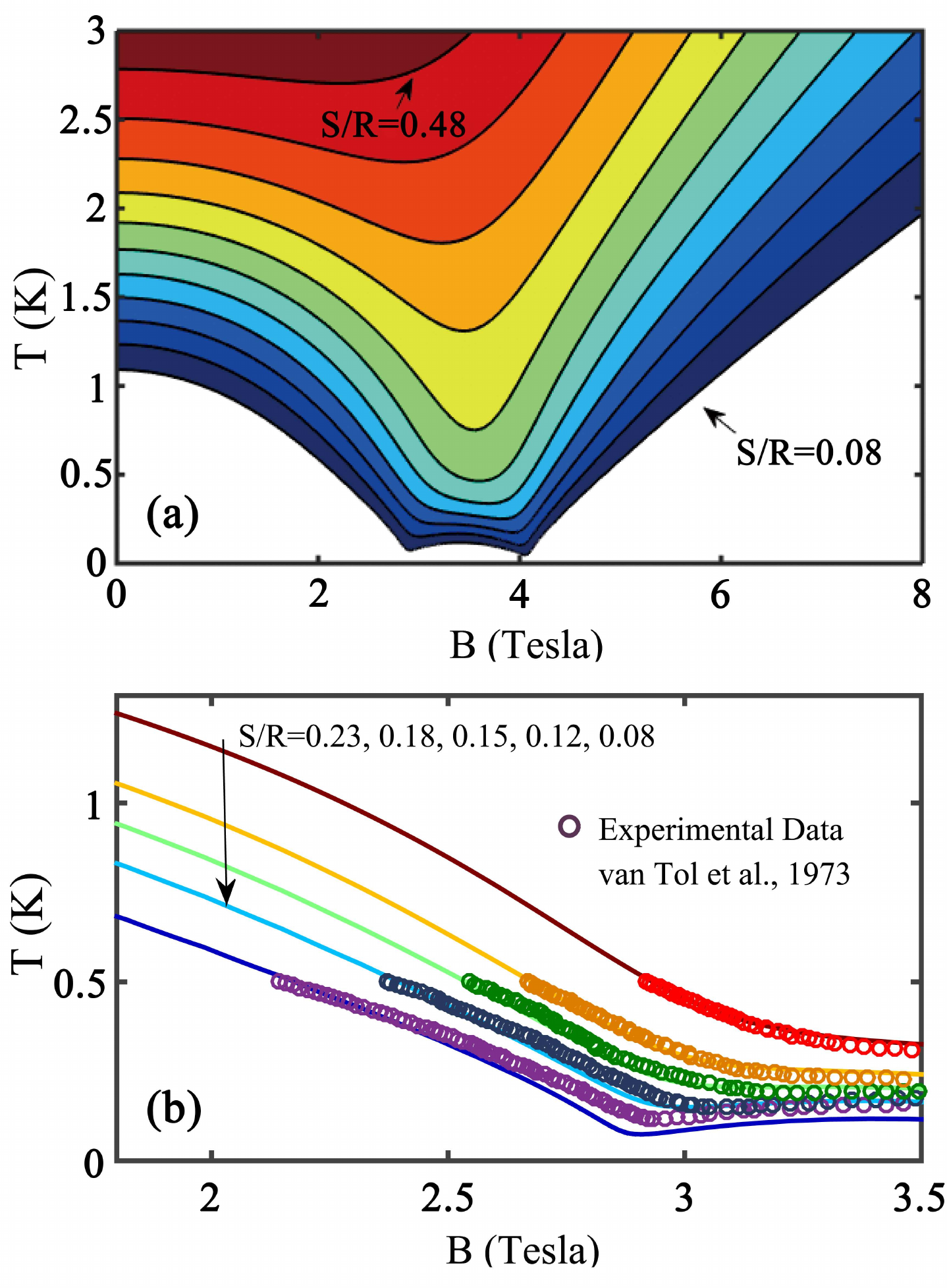}
  \caption{Numerically simulated and experimentally measured isentropes of CN. (a) The contour lines represent entropy per site $0.08\leq S/R \leq 0.48$ (bottom to top) with interval $\Delta S/R=0.04$,  where $R \simeq 8.314$ J $\cdot$ K$^{-1}$ $\cdot$  mol$^{-1}$ is the gas constant. (b) Comparisons to measured adiabatic isentropes of CN around the critical field $B_s \simeq 2.87$ T, the experimental data are taken from Refs. \onlinecite{Van-1973,Diederix-1979}.}
  \label{Fig:Isentrope}
\end{figure}

As early as 1968, magnetic refrigeration in spin chain material Cu(NO$_3$)$_2$ $\cdot$ 2.5H$_2$O has been experimentally observed, where a temperature decrease ranging from 1 to 58 mdeg was measured by increasing fields form 0.46 to 0.96 T \cite{Friedberg-1968}. Later on, van Tol \textit{et al.} \cite{Van-1973} explored the isentropic lines with fields in the range from 2 to 5 T, and observed two prominent dips with temperature as low as $\sim$ 100 mK (near two transition fields $B_c$ and $B_s$). This phenomenon was later attributed to large entropies near two fields, through ED calculations and theoretical analysis \cite{Diederix-1979}. However, it was unclear then that this enhanced MCE is due to quantum criticality at two transition fields, and a close comparison to experimental measured isentropes was absent, perhaps due to the lack of powerful manybody computation tools.

In this section, we employ the TTN simulations to explore the isentropes and magnetic Gr\"uneisen parameters, and revisit the early isentropic data in Ref. \onlinecite{Diederix-1979}. We reveal that there exists criticality-enhanced MCE near two field-induced QCPs. In Fig. \ref{Fig:MagCurves} the calculated magnetization at $T$=40 mK \cite{Footnote3}, where two QCPs, i.e., the plateau-closing field $B_c=2.87$ T and saturation fields $B_s=4.08$ T, are clearly shown. Between these two QCPs, there exists a continuous critical Luttinger liquid phase which hosts gapless magnetic excitations.

In Fig. \ref{Fig:Isentrope}(a), we plot the isentropic curves of various magnetic entropies (from $S/R =0.05$ to $0.5$). For curves with relatively large entropies ($0.2\leq S/R \leq 0.48$), the lowest temperature appears at around $B \simeq 3.5$ T, roughly located in the center of gapless region. However, with further lowering temperatures, we see that the broad dip eventually splits into two sharper dips in the isentropic curves, signalling two QCPs. Fig. \ref{Fig:Isentrope}(a) therefore manifests that in the both vicinities of two QCPs and in the quantum critical region, the thermal entropies are relatively large, which in turn results in criticality-enhanced MCEp.

Along each isentropic curve, one can read out the adiabatic temperature changes with varying magnetic fields. A quite distinct future of Fig. \ref{Fig:Isentrope}(a) for CN chain is that on both small and large field sides, one experiences large temperature changes by varying fields. This is in contrast to uniform Heisenberg model (see, for instance, Fig. 2 in Ref. \onlinecite{Wolf-2011} for spin chain material CuP), where only on large field (right) side of saturation QCP. For CuP chains, one can observe large MCE by decreasing from very large field to saturation point, while little temperature change is seen by increasing fields from 0 to saturation due to the presence of Luttinger liquid all along the magnetization curve. On the contrary, for the CN chain, the situation is different due to the existence of dimerization, which opens up a gap at low fields $< B_c$. This fact enables us to realize criticality-enhanced MCE for relatively small fields ($<$ 4 T), and one could even properly design a thermal cycling to make use enhance MCE around both low and high critical fields in one complete cooling process, with presumably much higher refrigeration efficiency than, say, uniform CuP chain \cite{Wolf-2011}.

In Fig. \ref{Fig:Isentrope}(b), we show the experimental data of isentropes (low-field region) from Refs. \onlinecite{Diederix-1979, Van-1973} and compare it to the simulated curves. From Fig. \ref{Fig:Isentrope}(b), we can see that, for isentropes with relatively large entropies (say, $S/R=0.23, 0.18, 0.15$), the fittings based on 1D model [Eq. (\ref{Eq-Hamiltonian})] are strikingly good; when the entropy decreases and the lowest temperature obtained in the adiabatic experiments reaches $\sim 100$ mK [see $S/R=0.08$ in Fig. \ref{Fig:Isentrope}(b)], slight deviation starts to show up in the vicinity of QCP with $B_c=2.87$ T. Such deviation may be ascibed to inter-chain interactions [see Eqs. (\ref{Eq:HalmitonianHC}, \ref{Eq:HalmitonianCTC})] since the magnitudes of $J_3$ and $J_m$ are both about $0.01$ meV ($\sim$ 100 mK). Nevertheless, the good agreements to adiabatic temperature changes evidences that CN indeed has criticality-enhanced MCE characterized by large temperature change even for moderate fields (say, from 0 to 3.5 T).

\begin{figure}[tbp]
  \includegraphics[angle=0,width=1.0\linewidth]{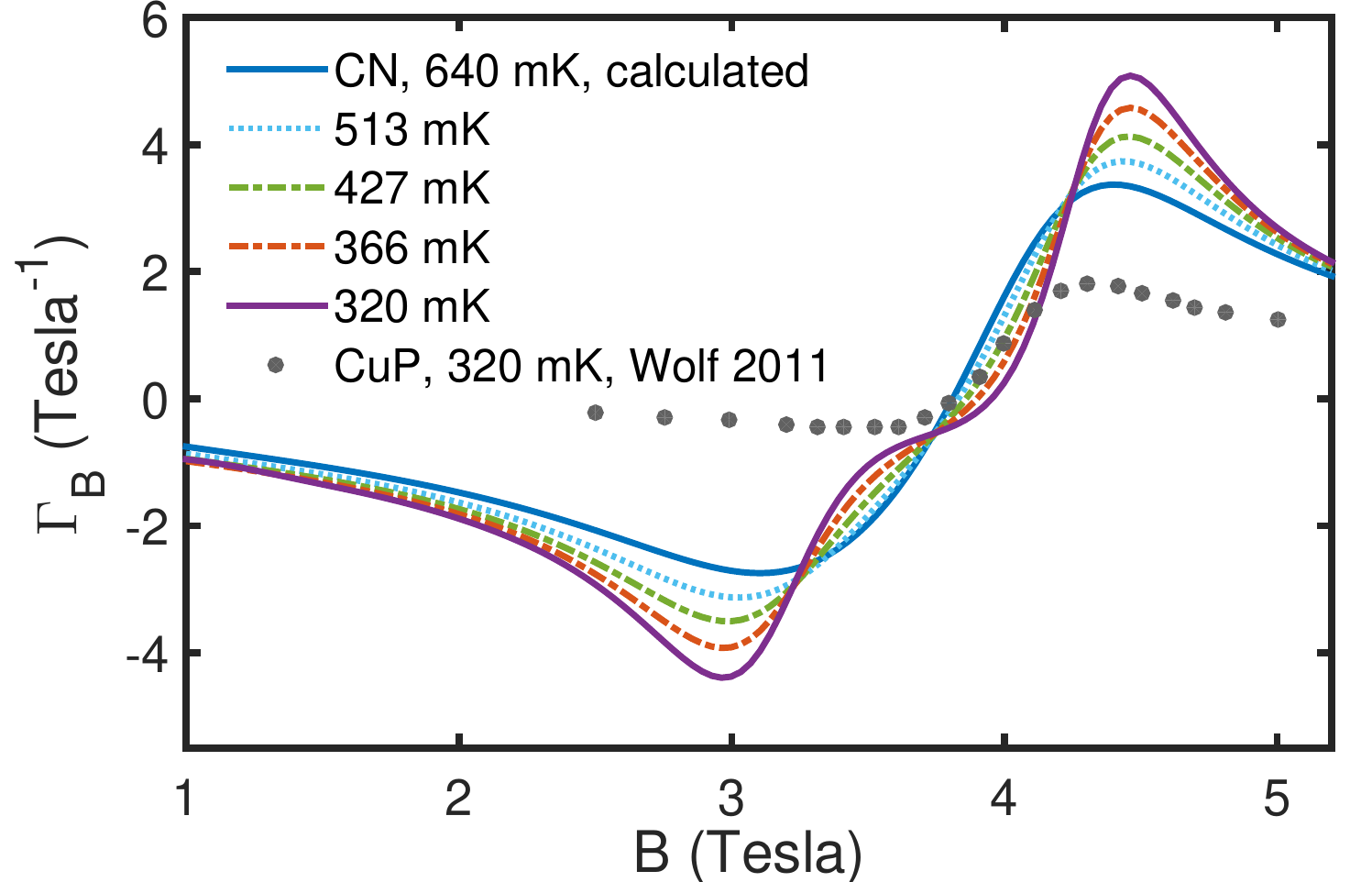}
  \caption{(Color online) Magnetic Gr\"ueisen parameter $\Gamma_B$ (definition in main text), which characterizes differentially the temperature change over an unit magnetic field change. The lines plotted, with different hight of peaks, correspond to different $\Gamma_B$ at various temperatures, which decrease from 640 mK to 320 mK (top to bottom).}
  \label{Fig:GP}
\end{figure}

Another important quantity measuring MCE property is the magnetic Gr\"uneisen parameter $\Gamma_B = \frac{1}{T} (\frac{\partial T}{\partial B})_s$, which is a differential characterization on the temperature change $\Delta T$ over small magnetic field variation $\Delta B$ in an adiabatic process. In the vicinity of QCPs, $\Gamma_B$ diverges as $T$ tends to zero, whose scaling behavior is intimately related to the quantum criticality \cite{Zhu-2003, Garst-2005}. In Fig. \ref{Fig:GP}, we show the calculated $\Gamma_B$ of CN, and also the measured $\Gamma_B$ of uniform spin-1/2 Heisenberg chain material CuP as a comparison (taken from Ref. \onlinecite{Wolf-2011}), from which it is seen that the CN chain has much larger $\Gamma_B$ around either one of its two QCPs, twice as large as that of CuP around the saturation field. The latter has been proposed as a perfect alternative for ordinary demagnetization refrigerant due to its wide operating range, large cooling power, and high efficiency \cite{Wolf-2011}. Our simulations here shows that the dimerized spin chain CN studied in the present work has even more promising potential as quantum critical coolant, not only because it has two sharp dips at suitable fields (Fig. \ref{Fig:Isentrope}), one at $B_c=2.87$ and one at $B_s=4.08$ T, but also due to large temperature change in response to the same field variation as revealed by calculated $\Gamma_B$ shown in Fig. \ref{Fig:GP}.

\section{Conclusion and Outlook}
\label{Sec:Summary}
In this paper, we generalize the linearized tensor renormalization group (LTRG) method to a bilayer form which is essentially equivalent to the well-established TMRG method, and employ this cutting-edge TTN method to accurately study the thermodynamics of a 1D dimerized spin chain material copper nitrate. We calculate and fit the experimental data of specific heat, magnetic susceptibility, and magnetization curves, some of which are measured experimentally in the present work. The previous discrepancy in coupling constants determined from different experiments are resolved, and we conclude that the set of parameters $J=5.13$ K, $\alpha=0.23(1)$, $g_{\parallel}=2.31$, and $g_{\perp}=2.14$ yielded from fitting thermal properties, is actually in remarkable consistency with those determined from INS experiments. In addition, based on electron density distribution pattern, we have for the first time visualized the spin-chain exchange path in CN, through ab initio calculations. On the grounds of these calculations, and also according to INS experiments, we obtain a 3D spin model describing magnetic properties of CN below about 160 mK.

With this set of parameters characterizing the magnetic properties of spin chain model material CN, we uncover, though accurate TTN simulations, that there exists criticality-enhanced large MCE near two quantum phase transition points, even at very low temperatures. Judged from the quantum anomaly in low-temperature isentropic curves and the extraordinarily good agreements to experimentally measured isentropes, as well as the large peaks/dips in magnetic Gr\"uneisen parameters, we propose that CN is a very promising quantum critical coolant with significant temperature changes in response to magnetic field variations of moderate values.

There are still a number of interesting questions deserving further discussions, on both experimental and theoretical sides. To name a few, the direct experimental measurement of adiabatic temperature change for larger field ranges, instead of the rather limited field range between 2 to 4 T in previous experiments, is important to verify our prediction of CN as a promising quantum critical coolant. In addition, the performance characteristics such as operation temperature range, cooling power, and efficiency, are also in due to be investigated. Another important ingredient missing in the present work is the effect of inter chain couplings, whose Hamiltonian has been given in Eqs. (\ref{Eq:HalmitonianHC}, \ref{Eq:HalmitonianCTC}). The inter-chain couplings could be of importance since the coolant is supposed to work in a circumstance with lowest temperature $T<$ 100 mK, an energy scale comparable to that of inter-chain exchange constant in CN.

\begin{acknowledgments}
WL is indebted to Tao Xiang, Shi-Ju Ran, Ji-Ze Zhao, Hong-Liang Shi, Fei Ye, and Peijie Sun for useful discussions. WL also acknowledges Shou-Shu Gong for nicely providing (part of) Fig. \ref{Fig:TTN}. This work was supported by the National Natural Science Foundation of China (Grant Nos. 11274033, 11474015, 11504014, and 61227902), the Research Fund for the Doctoral Program of Higher Education of China (Grant No. 20131102130005), and the Beijing Key Discipline Foundation of Condensed Matter Physics.
\end{acknowledgments}
\begin{appendix}
\setcounter{equation}{0}
\setcounter{figure}{0}
\setcounter{table}{0}
\makeatletter
\renewcommand{\theequation}{A\arabic{equation}}
\renewcommand{\thefigure}{A\arabic{figure}}

\section{Bilayer thermal tensor network algorithm LTRG++, hidden Matrix Product State, and comparisons to TMRG}
\label{App:LTRG++}

\begin{figure}[tbp]
  \includegraphics[angle=0,width=0.8\linewidth]{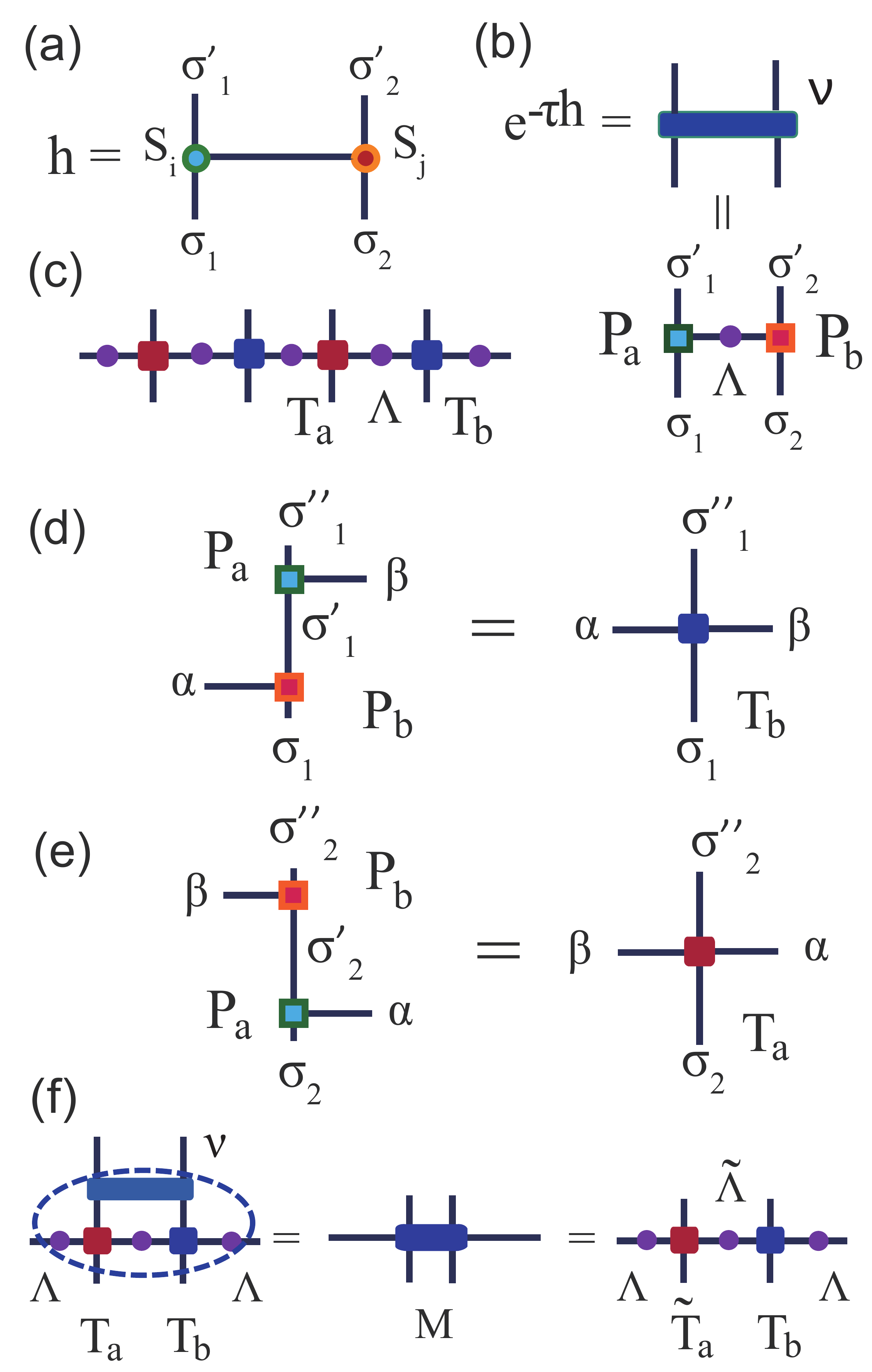}
  \caption{(Color online) (a) A local coupling term $ \vec{S}_i \cdot \vec{S}_j$, where $\vec{S} = \{ S^x, S^y, S^z \}$ is a vector spin operator. (b) Elementary tensor $\nu = \exp{(-\tau h)}$ can be decomposed into two rank-three tensors $P_a$ and $P_{b}$. (c) Matrix product operator representation of density matrix of spin chain. (d,e) Rank-four tensors $T_{a(b)}$ are obtained by recombining $P_{b}$ and $P_{a}$, which interconnects via the geometric (horizontal) bonds and form an MPO. (f) Local projection and truncation scheme.}
  \label{Fig:LocalTens}
\end{figure}

\begin{figure}[tbp]
  \includegraphics[angle=0,width=1\linewidth]{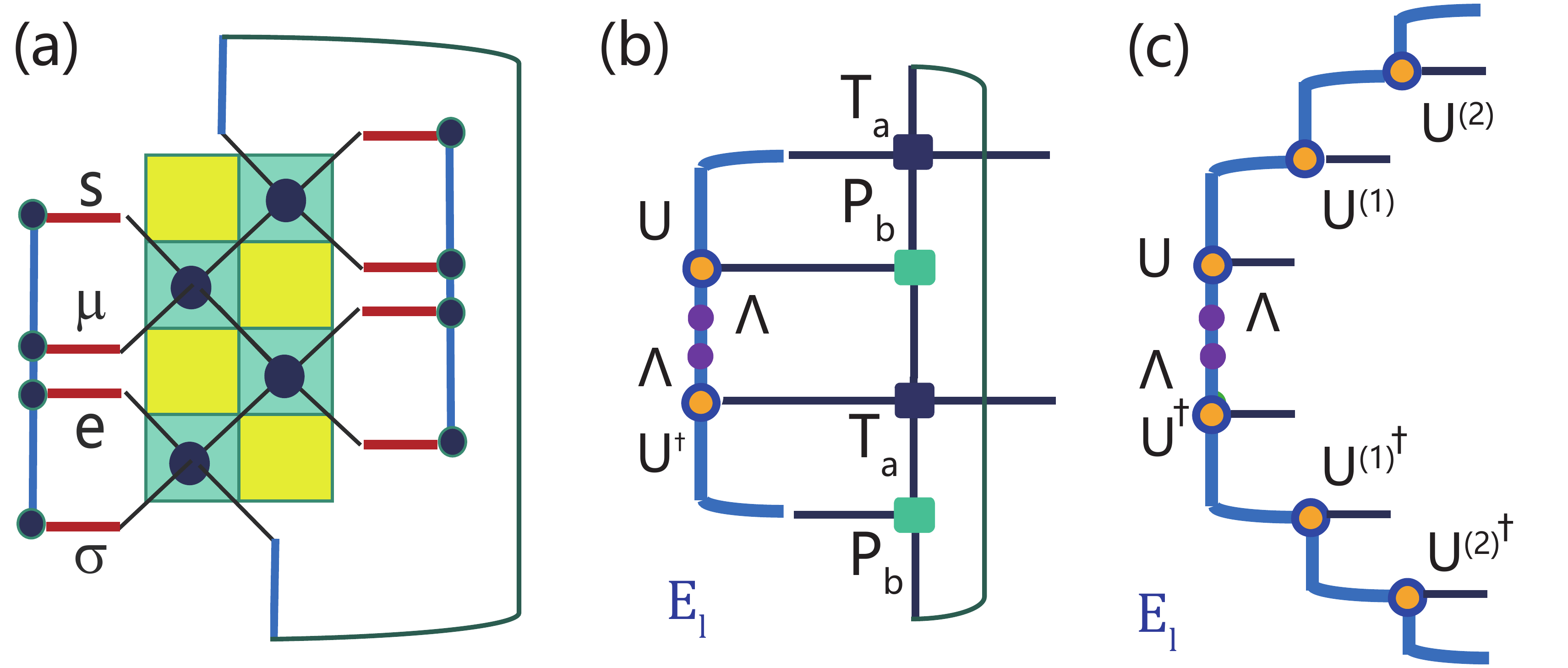}
  \caption{(Color online) (a) Traditional TMRG algorithm, four blocks $s, \mu, e, \sigma$ constitutes a symmetric construction. (b) Bilayer LTRG++ algorithm for contracting thermal tensor network, $E_l \simeq U \Lambda^2 U^{\dagger}$ is the dominant left eigenvector of the transfer matrix, where $\Lambda$ is a diagonal matrix obtained in LTRG++. (c) Hidden matrix product state representation of $E_l$ in the LTRG++, revealed explicitly by expanding $E_l$ with $\{U^{(i)}\}$ series [and also $(U^{(i)})^{\dagger}$].}
  \label{Fig:equivalence}
\end{figure}

In this appendix, we provide more technical details of linearized tensor renormalization group (LTRG) and LTRG++ algorithms and relate them to the well-established transfer-matrix renormalization group (TMRG) approach. In Fig. \ref{Fig:LocalTens}(a), we firstly specify the local two-site Hamiltonian as $h_{i,j} = J_{i,j} \vec{S}_i \cdot \vec{S}_j$ between spin-1/2's. Such local spin-spin coupling term can be seen in the AHAFC model Eq. (\ref{Eq-Hamiltonian}) of main text. Given the local tensor $\nu_{\sigma_1, \sigma_2, \sigma'_1, \sigma'_2} = e^{-\tau h_{i,j}}$ as in Figs. \ref{Fig:LTRG++}(b), one decomposes $\nu$ into $P_{a(b)}$ and then recombine them in a way [Figs. \ref{Fig:LTRG++}(d,e)] to form the rank-four tensor $[T_{b(a)}]_{\alpha(\beta), \beta(\alpha), \sigma_{1(2)}, \sigma''_{1(2)}}=\sum_{\sigma'_{1(2)}} [P_{b(a)}]_{\alpha(\beta),\sigma_{1(2)},\sigma'_{1(2)}} \cdot [P_{a(b)}]_{\beta(\alpha),\sigma'_{1(2)},\sigma''_{1(2)}}$.
$T_{a,b}$'s are interconnected with each other via geometric bonds $\alpha, \beta$ and constitute an matrix product operator (MPO) of initial bond dimension $D=4$ [Fig. \ref{Fig:LocalTens}(c)].

After the initialization, we need to evolve the MPO along Trotter direction with the local projection and truncation scheme shown in Fig. \ref{Fig:LocalTens}(f). In either LTRG or LTRG++, we absorb the evolving operator $\nu$ to local tensors $T_a, T_b$ in the MPO, as well as two bond weights $\Lambda$, and arrive at the $M$ tensor of rank six. Then one reshape $M$ into a matrix and decompose it to obtain two updated tensors $\tilde{T_a}, \tilde{T_b}$, and new bond weight $\tilde{\Lambda}$ (find more details in Ref. \onlinecite{LTRG}). Note that since the so obtained MPO is quasi-canonical due to that every single evolving operator $\nu$ is close to identity.

In order to calculate the free energy per site at any specific temperature, following Eq. (\ref{Eq:LTRG++}) in the main text, we need to collect the normalization factors $\kappa_i^{a(b)}$ (obtained by normalizing $\Lambda$ after each projection process) and the dominant eigenvalue $\theta_{\rm{max}}$ of ladder transfer matrix shown in Fig. \ref{Fig:LTRG++}(f) of main text. Notice that a proper combination of $\kappa_i$'s and $\lambda_{\rm{max}}$, i.e., $ \theta_{\rm{max}} = (\Pi_{i=1}^{n} \kappa^a_i \kappa_i^b) \cdot \lambda_{\rm{max}}$, is nothing but an accurate estimate of dominant eigenvalue of vertical transfer matrix [see Fig. \ref{Fig:TTN}(b) of main text] one pursues in the TMRG algorithm.

To be specific, we compare in Fig. \ref{Fig:equivalence} the LTRG++ algorithm with traditional TMRG. As shown in Fig. \ref{Fig:equivalence}(a), TMRG exploits the ``s-$\mu$-e-$\sigma$" construction, adding two blocks ($\mu$ and $\sigma$) per iteration, and truncate enlarged system (environment) block $s$-$\mu$ ($e$-$\sigma$) with the aid of dominant eigenvectors of transfer matrix. Note that the transfer matrix is non-Hermitian and have two sets (left and right) of eigenvectors. Take the dominant left eigenvector as an example: one can construct the reduced density matrix of $s$-$\mu$ (or $e$-$\sigma$), and retain the largest $m$ states (White's rule) to reduce the enlarged direct-product space $s$-$\mu$ ($e$-$\sigma$), executing the truncation of left indices of the transfer matrix. The procedure is similar for the truncation of right indices, with the help of dominant \textit{right} eigenvector.

\begin{figure}[tbp]
  \includegraphics[angle=0,width=1\linewidth]{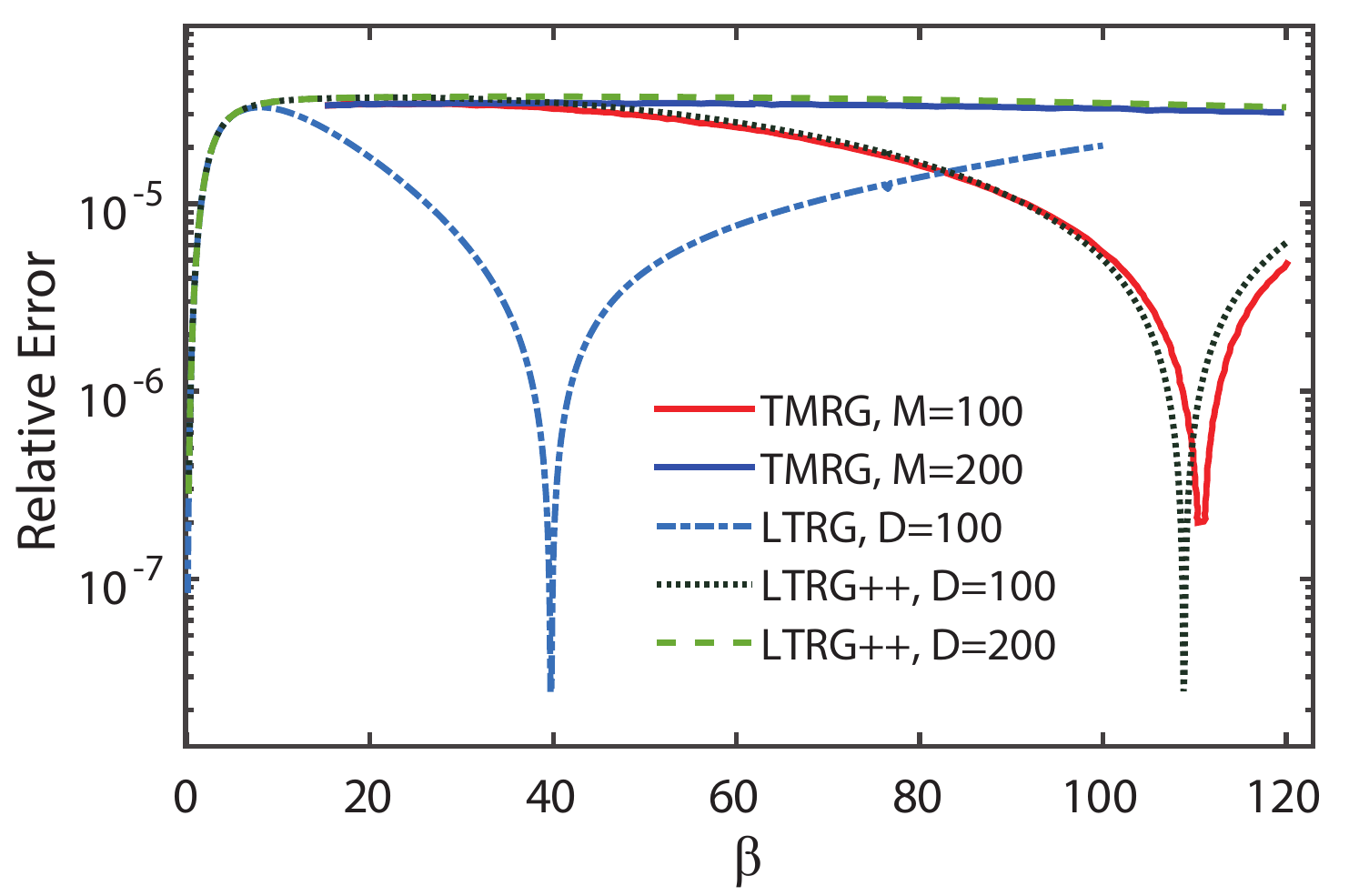}
  \caption{(Color online) Relative errors of free energy per site in LTRG, LTRG++, and TMRG, for the exactly soluble model 1D free fermion chain. $M$ and $D$ are the number of retained states during the process of projection and truncation in LTRG/LTRG++ and TMRG, respectively.}
  \label{Fig:comparison}
\end{figure}

At first sight, this algorithm, firstly proposed in the language of traditional DMRG \cite{TMRG}, is quite different from the bilayer LTRG++ algorithm introduced in section \ref{SubSec:LTRG++} of main text. The latter is developed in the language of tensor networks, and it contracts the tensor networks along Trotter direction, seemingly different from the former. However, in fact, these two algorithms are, despite some minor technical details, essentially equivalent. In Figs. \ref{Fig:equivalence}(b,c), we show that there exists a hidden matrix product state (MPS) in LTRG++ algorithm, which is nothing but the dominant eigenvector of vertical transfer matrix in TMRG. In Fig. \ref{Fig:equivalence}(b), take the left eigenvector $E_l$ as an example, it is shown that $E_l \simeq U \Lambda^2 U^{\dagger}$ where $\Lambda$ is a diagonal matrix storing the Schmidt coefficients, obtained from singular value decomposition on the specific bond \cite{LTRG}. Given that, we can further expand $E_l$ in terms of the series of $U^{(i)}$ and $(U^{(i)})^{\dagger}$, the truncation matrices generated in each projection step (not stored during the process of computations, though), and arrive at Fig. \ref{Fig:equivalence}(c) which reveals explicitly the hidden MPS representation of dominant eigenvector of transfer matrix.

To confirm the above argument, in Fig. \ref{Fig:comparison} we compare numerically the accuracy of single-layer LTRG, TMRG and bilayer LTRG++ results for the exactly soluble XY spin chain model. Note that there exists a point where the truncation error happens to cancel the Trotterp error completely, leading to a very steep dip in Fig. \ref{Fig:comparison}: For small $\beta$ (high temperatures regime) Trotter error is dominant, while for large $\beta$ (low temperature regime) truncation error takes the place. Since the Trotter error accumulates linearly with $\beta$ and is independent of the specific truncation schemes, the position of the cancellation point thus reveals how fast truncation errors accumulate and plays the role as an indicator of ``goodness" of the truncation scheme. In principle, the further the cancellation point dwells on $\beta$ axis, the better the truncation scheme is. Therefore, according to Fig. \ref{Fig:comparison}, bilayer LTRG++ has significant improvement compared to the single-layer LTRG, and bears practically the same accuracy as TMRG. This observation strongly supports our argument above that LTRG++ is in essence equivalent to TMRG.

To summarize, although it seems that LTRG++ and TMRG are following reversed orders in contracting the TTN (Trotter or spatial direction first), a more careful analysis, though, reveals that the truncation matrices in LTRG++ constitute a hidden MPS and these two methods are essentially equivalent. This can be attributed to the fact that a full contraction of TTN must consider both directions in equal footing and the superficial order of contractions does not really matter. In fact, these two renormalization group (RG) algorithms, and potentially other RG based thermal algorithms, can be unified in the framework of TTN simulations, more details on this topic will be published elsewhere.

\section{Sample Preparation}
\label{App:Sample}
\begin{figure}[tbp]
  \includegraphics[angle=0,width=0.95\linewidth]{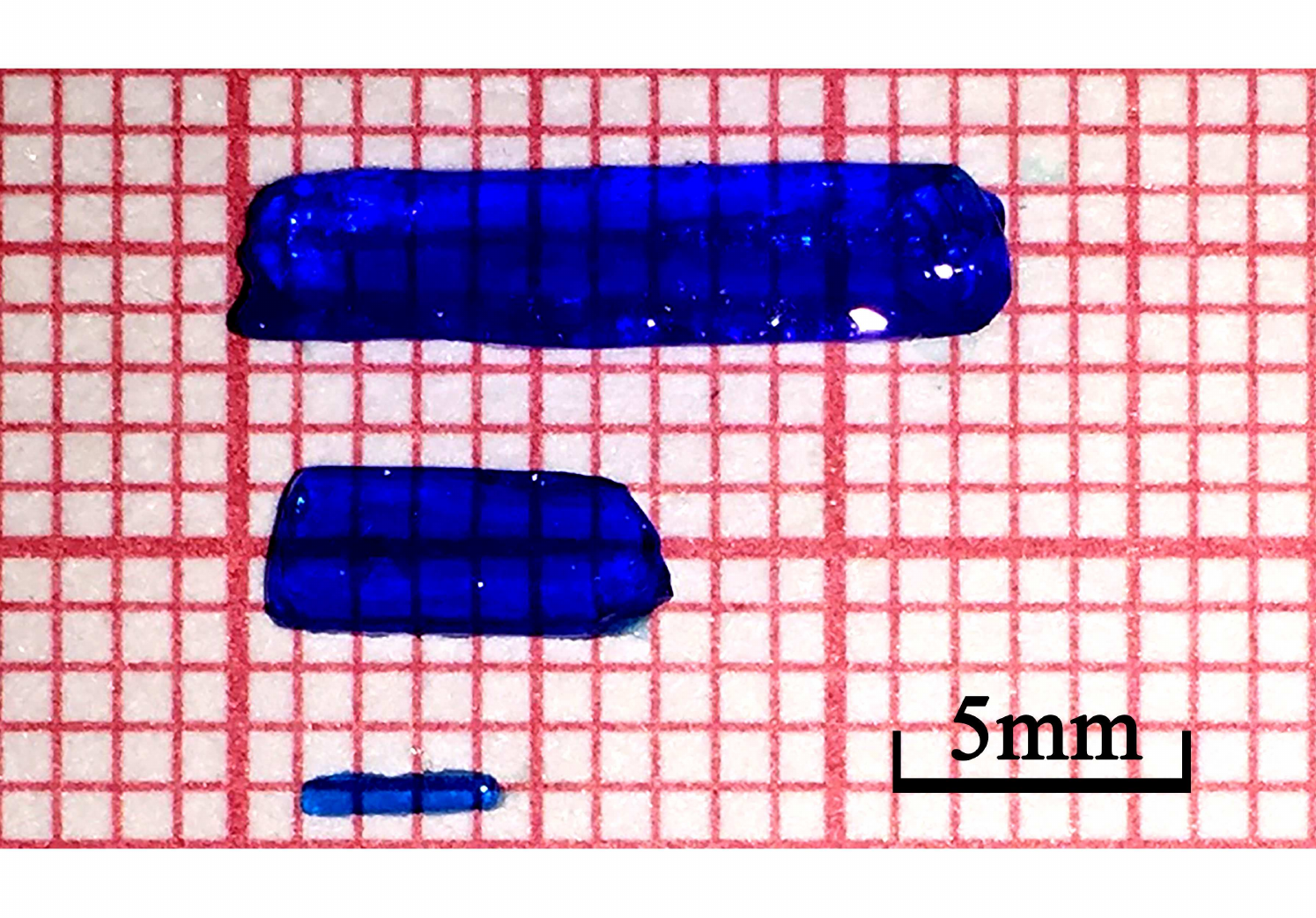}
  \caption{(Color online) Examples of synthesized single-crystal CN specimens, which are transparent and with needle-like shapes. The needle axis is parallel to crystal $b$ axis, and the lengths of samples range from a few mm to more than 1 cm.}
  \label{Fig:sample}
\end{figure}

The Cu(NO$_3$)$_2$ $\cdot$ 2.5H$_2$O single crystals are obtained by cooling the hot saturated water solution of copper nitrate (CN) down to low temperatures. The solution was heated to increase the concentration, but the highest temperature should be below 75 $^\circ$C to prevent copper nitrate from decomposition \cite{Friedberg-1968}. In practice, we heat the hot solutions to 75 $^\circ$C, and then transfer it directly to a cooler container ($< 25 ^\circ$C) to facilitate crystal seed formation. The temperature of latter controls the final single-crystal size of the specimen. Sequentially, the container is put into the 60 $^\circ$C environment, which is slowly cooled down to 35 $^\circ$C. The grown single crystals have quite large system sizes, ranging from several milimeters to one or two centimeters (see Fig. \ref{Fig:sample}), and are in needle shapes with the long edge right along the crystal $b$ axis.p Note during the process of sample preparation, the CN solution should be kept away from organic materials or solution.

In order to ensure the purity of Cu(NO$_3$)$_2$ $\cdot$ 2.5H$_2$O in the specimen (i.e., to remove superfluous water and other possible CN hydrates), the sample is heat to 45 $^\circ$C for 10 min everytime before measurements \cite{Morozov-2003}. The data in Figs. \ref{Fig:Sus}, \ref{Fig:MagCurves}, including the isothermal magnetization curves and the zero-field susceptibility are measured in the high-precision SQUID device by scanning fields and temperatures, respectively.

\end{appendix}


\begin{thebibliography}{99}
% quantum spin liquid & other exotic quantum magnetism
\bibitem{Balents-2010} L. Balents, \textit{Spin liquids in frustrated magnets}, Nature \textbf{464}, 7286 (2010).
\bibitem{Yan-2011} S. Yan, D. A. Huse, and S. R. White, \textit{Spin-liquid ground state of the S = 1/2 Kagome Heisenberg antiferromagnet}, Science \textbf{332}, 1173 (2011).
\bibitem{Rahmani-2014} A. Rahmani, A. E. Feiguin, and C. D. Batista, \textit{Anyonic Liquids in Nearly Saturated Spin Chains}, Phys. Rev. Lett. \textbf{113}, 267201 (2014).
\bibitem{Zapf-2014} V. Zapf, M. Jaime, and C. D. Batista, \textit{Bose-Einstein condensation in quantum magnets}, Rev. Mod. Phys. \textbf{86}, 563 (2014).
\bibitem{Kikuchi-2005} H. Kikuchi, et al., \textit{Experimental Observation of the 1/3 Magnetization Plateau in the Diamond-Chain Compound Cu$_3$(CO$_3$)$_2$(OH)$_2$}, Phys. Rev. Lett. \textbf{94}, 227201 (2005).
\bibitem{Han-2012} T.-H. Han, et al., \textit{Fractionalized excitations in the spin-liquid state of a kagome-lattice antiferromagnet}, Nature \textbf{492}, 406 (2012).

% copper nitrate
\bibitem{Berger-1963} L. Berger, S. A. Friedberg, and J. T. Chriempf, \textit{Magnetic Susceptibility of Cu(NO$_3$)$_2$ $\cdot$ 2.5H$_2$O at Low Temperature}, Phys. Rev. \textbf{132}, 1057-1061 (1963).
\bibitem{Friedberg-1968} S. A. Friedberg and C. A. Raquet, \textit{The Heat Capacity of Cu(NO$_3$)$_2$ $\cdot$ 2.5H$_2$O at Low Temperatures}, Journal of Applied Physics \textbf{39}, 1132 (1968).
\bibitem{Garaj-1968} J. Garaj, \textit{Crystal Structure of Cu(NO$_3$)$_2$ $\cdot$ 2.5H$_2$O}, Acta Chem. Scand. \textbf{22}, 1710 (1968).
\bibitem{Amaya-1969} K. Amaya, et al, \textit{Paramagnetic relaxation and adiabatic cooling in Cu(NO$_3$)$_2$ $\cdot$ 2.5H$_2$O}, Phys. Lett. A. \textbf{28}, 732(1969).
\bibitem{Myers-1969} B. E. Myers, L. Berger, and S. A. Friedberg, \textit{Low-Temperature Magnetization of Cu(NO$_3$)$_2$ $\cdot$ 2.5H$_2$O}, J. Appl. Phys. \textbf{40}, 1149 (1969).
\bibitem{Morosin-1970} B. Morosin, \textit{The crystal structure of Cu(NO$_3$)$_2$ $\cdot$ 2.5H$_2$O}, Acta Crystallographica Section B: Structural Crystallography and Crystal Chemistry \textbf{26}, 1203-1208 (1970).
\bibitem{Van-1971} M. W. van Tol, L. S. J. M. Henkens, and N. J. Poulis, \textit{High-Field Magnetic Phase Transition in Cu(NO$_3$)$_2$ $\cdot$ 2.5H$_2$O}, Phys. Rev. Lett. \textbf{27}, 739 (1971).
\bibitem{Van-1972} M. W. van Tol, H. M. C. Eijkelhof and A. J. Van Duyneveldt, \textit{Energy levels and relaxation effects in Cu(NO$_3$)$_2$ $\cdot$ 2.5H$_2$O}, Physica \textbf{60}, 223 (1972).
%\bibitem{Haseda-1970} T. Haseda, Y. Tokunaga, Y. Kuramitsu, K. Amaya and S. Sakatsume, to be published in  \textit{Proceedings of the Twelfth International Conference on Low Temperature Physics}, Kyoto (1970)
\bibitem{Van-1973} M. W. Van Tol, J. P. Groen and N. J. Poulis, \textit{Specific heat and NMR of Cu(NO$_3$)$_2$ $\cdot$ 2.5H$_2$O at the high-field phase transition}, Physica \textbf{64}, 363 (1973).
\bibitem{Amaya-1977} K. Amaya and N. Yamashita, \textit{Adiabatic Magnetization Cooling in Cu(NO$_3$)$_2$ $\cdot$ 2.5H$_2$O by Pulsed Magnetic Field}, J. Phys. Soc. Jpn. \textbf{42}, 24 (1977).
\bibitem{Diederix-1977} K. M. Diederix, J. P. Groen and N. J. Poulis, \textit{A study of the high field phase transition in Cu(NO$_3$)$_2$ $\cdot$ 2.5H$_2$O}, Physica \textbf{86}, 1151 (1977).
\bibitem{Diederix-1978a} K. M. Diederix, et al., \textit{An experimental study on the magnetic properties of the singlet ground-state system in Cu(NO$_3$)$_2$ $\cdot$ 2.5H$_2$O: I. Short-range ordered state}, Physica \textbf{93B}, 99 (1978).
\bibitem{Diederix-1978b} K. M. Diederix, et al., \textit{An experimental study on the magnetic properties of the singlet ground-state system in Cu(NO$_3$)$_2$ $\cdot$ 2.5H$_2$O: II. The long-range ordered state}, Physica \textbf{94B}, 9 (1978).
\bibitem{Diederix-1979} K. M. Diederix, et al., \textit{Theoretical and experimental study of the magnetic properties of the singlet-ground-state system Cu(NO$_3$)$_2$ $\cdot$ 2.5H$_2$O: An alternating linear Heisenberg antiferromagnet}, Phys. Rev. B. \textbf{19}, 420 (1979).
\bibitem{Eckert-1979} J. Eckert, et al., \textit{Magnetic ordering in Cu(NO$_3$)$_2$ $\cdot$ 2.5D$_2$O,} Phys. Rev. B. \textbf{20}, 4596 (1979).
\bibitem{Bonner-1983} J. C. Bonner, et al., \textit{Alternating linear-chain antiferromagnetism in copper nitrate Cu(NO$_3$)$_2$ $\cdot$ 2.5H$_2$O}, Phys. Rev. B. \textbf{27}, 248 (1983).
\bibitem{Xu-1999} Guangyong Xu, C. Broholm, Daniel H. Reich, and M. A. Adams, \textit{Triplet Waves in a Quantum Spin Liquid}, Phys. Rev. Lett. \textbf{84}, 4465 (1999).
\bibitem{Taylor-1986} T. J. Taylor, D. Dollimore and G. A. Gamlen, \textit{Deaquation and denitration studies on copper nitrate trihydrate}, Thermochimica. Acta. \textbf{103}, 333-340 (1986).
\bibitem{Tennant-2003} D. A. Tennant, et al., \textit{Neutron scattering study of two-magnon states in the quantum magnet copper nitrate}, Phys. Rev. B. \textbf{67}, 054414 (2003).
\bibitem{Morozov-2003} I.V. Morozov, K.O. Znamenkov, Yu.M. Korenev, O.A. Shlyakhtin, Thermal decomposition of Cu(NO$_3$)$_2$ $\cdot$ 3H$_2$O at reduced pressures, Thermochimica. Acta. \textbf{403}, 173-179 (2003).
\bibitem{Stone-2014} M. B. Stone, et al., \textit{Magnons and continua in a magnetized and dimerized spin-1/2 chai}n, Phys. Rev. B. \textbf{90}, 094419 (2014).
\bibitem{Willenberg-2015} B. Willenberg, et al., \textit{Luttinger liquid behavior in the alternating spin-chain system copper nitrate}, Phys. Rev. B. \textbf{91}, 060407(R) (2015).

% MCE
\bibitem{Warburg-1881} E. Warburg, \textit{Magnetische Untersuchungen. Ueber einige Wirkungen der Co\"ercitivkraft.}, Ann. Phys. (Leipzig) \textbf{13}, 141 (1881).
\bibitem{Weiss-1917} P. Weiss and A. Piccard, \textit{Le ph\'enome\'ne magne\'tocalorique}, J. Phys. (\textit{Paris}) \textbf{7}, 103 (1917).
\bibitem{Smith-2013} A. Smith, \textit{Who discovered the magnetocaloric effect?} Eur. Phys. J. H \textbf{38}, 507 (2013).
\bibitem{Gschneidner-1997} V. K. Pecharsky, K. A. Gschneidner, \textit{Giant Magnetocaloric Effect in Gd$_{5}$(Si$_{2}$Ge$_{2}$)}, Phys. Rev. Lett. \textbf{78}, 4494 (1997).
\bibitem{Smith-2012} A. Smith, C. R.H. Bahl, R. Bj$\phi$rk, K. Engelbrecht, K. K. Nielsen, and N. Pryds, \textit{Materials Challenges for High Performance Magnetocaloric Refrigeration Devices}, Adv. Energy Mater. \textbf{2}, 1288 (2012).
\bibitem{Brown-1976} G. V. Brown, \textit{Magnetic heat pumping near room temperature}, Journal of Alloys and Compounds \textbf{47}, 3673 (1976).
\bibitem{Pecharsky-1997} V. K. Pecharsky, K. A. Gscdneidner, \textit{Giant magnetocaloric effect in Gd$_5$(Si$_2$Ge$_2$)}, Phys. Rev. Lett. \textbf{78}, 4494 (1997).
\bibitem{Hagmann-1995} C. Hagmann, P. L. Richards, \textit{Adiabatic demagnetization refreigeratiors for small laboratory experiments and space astronomy}, Cryogenics \textbf{35}, 303 (1995).
\bibitem{Zimm-2003} C. B. Zimm, A. Sternberg, A. G. Jastrab, A. M. Boeder, L. M. Lawton, J. J. Chell, \textit{Rotating bed magnetic refrigeration apparatus}, US Patent 6.526.759.4 (2003).
\bibitem{Giauque-1927} W. F. Giauque, \textit{A thermodynamic treatment of certain magnetic effects, A proposed method of producing temperatures considerably below 18$^{\circ}$ absolute}, Journal of American Chemical Society \textbf{49}, 1864 (1927).
\bibitem{Giauque-1933} W. F. Giauque, D. P. MacDougall, \textit{Attainment of Temperatures Below 1$^{\circ}$ Absolute by Demagnetization of Gd$_2$(SO$_4$)$_3$ $\cdot$ 8H$_2$O}, Phys. Rev. \textbf{43}, 786 (1933).

% MCE & quantum criticality
\bibitem{Sharples-2014} J. W. Sharples, D. Collison, E. J. L. Mclnnes, J. Schnack, E. Palacios, and M. Evangelisti, \textit{Quantum Signatures of a molecular nanomagnet in a direct magnetocaloric measurements}, Nature Commun. \textbf{5} 5321 (2014).
\bibitem{Rost-2009} A. W. Rost, et al., \textit{Entropy Landscape of Phase Formation Associated with Quantum Criticality in Sr$_3$Ru$_2$O$_7$}, Science \textbf{325}, 1360 (2009).
\bibitem{Ryll-2014} H. Ryll, et. al, \textit{Magnetic entropy landscape and Gr\"uneisen parameter of a quantum spin ladder}, Phys. Rev. B \textbf{89}, 144416 (2014).
\bibitem{Zhu-2003} L. Zhu, M. Garst, A. Rosch, and Q. Si, \textit{Universally Diverging Gr\"uneisen Parameter and the Magnetocaloric Effect Close to Quantum Critical Point}, Phys. Rev. Lett. \textbf{91}, 066404 (2003).
\bibitem{Garst-2005} M. Garst and A. Rosch, \textit{Sign change of the Gr\"uneisen parameter and magnetocaloric effect near quantum critical points}, Phys. Rev. B \textbf{72}, 205129 (2005).
\bibitem{Lucia} L. G\'alisov\'a, \textit{Magnetocaloric effect in the spin-1/2 Ising-Heisenberg diamond chain with the four-spin interaction}, Condens. Matter Phys. \textbf{17}, 13001 (2014).
\bibitem{Zhitomirsky-2004} M. E. Zhitomirsky, A. Honecker, \textit{Magnetocaloric effect in one-dimensional antiferromagnets}, J. Stat. Mech.: Theor. Exp. P07012 (2004).
\bibitem{Honecker-2009} A. Honecker, S. Wessel, \textit{Magnetocaloric effect in quantum spin-s chains}, Condens. Matter Phys. \textbf{12}, 399 (2009).
\bibitem{Lang-2010} M. Lang, et. al., \textit{Large Magnetocaloric Effect at the Saturation Field of an S=1/2 Antiferromagnetic Heisenberg Chain}, J. Low Temp. Phys. \textbf{159}, 88 (2010).
\bibitem{Wolf-2011} B. Wolf, et. al., \textit{Magnetocaloric effect and magnetic cooling near a field-induced quantum-critical point}, Proc. Natl. Acad. Sci. USA \textbf{108}, 6862 (2011).
%\bibitem{Ding-2016} L. J. Ding, Y. Zhong, S. W. Fan and L. Y. Zhu, \textit{The magnetocaloric effect with critical behavior of a periodic Anderson-like organic polymer}, Phys. Chem. Chem. Phys. \textbf{18}, 510 (2016).

% ThTN methods
\bibitem{TMRG} R.J. Bursilly, T. Xiang and G.A. Gehring, \textit{The density matrix renormalization group for a quantum spin chain at non-zero temperature}, J. Phys.: Condes. Matter \textbf{8}, L583 (1996); X. Wang and T. Xiang, \textit{Transfer-matrix density-matrix renormalization-group theory for thermodynamics of one-dimensional quantum systems}, Phys. Rev. B \textbf{56}, 5061 (1997); T. Xiang, \textit{Thermodynamics of quantum Heisenberg spin chain}, Phys. Rev. B \textbf{58} 9142 (1998).
\bibitem{DMRG-19921993} S.R. White, \textit{Density Matrix Formulation for Quantum Renormalization Groups}, Phys. Rev. Lett. \textbf{69}, 2863 (1992); S.R. White, \textit{Density-matrix algorithms for quantum renormalization groups}, Phys. Rev. B \textbf{48} 10345 (1993).
\bibitem{LTRG} W. Li, S. J. Ran, S. S. Gong, Y. Zhao, B. Xi, F. Ye, and G. Su, \textit{Linearized Tensor Renormalization Group Algorithm for the Calculation of Thermodynamic Properties of Quantum Lattice Models}, Phys. Rev. Lett. \textbf{106}, 127202 (2011).
\bibitem{Suzuki-1976} M. Suzuki, \textit{Relationship between d-Dimensional Quantal Spin Systems and (d+1)-Dimensional Ising Systems: Equivalence, Critical Exponents and Systematic Approximants of the Partition Function and Spin Correlations}, Prog. Theor. Phys. \textbf{56}, 1454 (1976).

\bibitem{Wittekoek-1968} S. Wittekoek and N. J. Poulis, \textit{Proton Magnetic Resonance Study of Magnetic Ordering in Two Cupric Salts}, Journal of Applied Physics \textbf{39}, 1017 (1968).
\bibitem{Sakai-1995} T. Sakai, \textit{Phase Transition of S=1/2 Bond-Alternating Chain in a Magnetic Field}, J. Phys. Soc. Jpn. \textbf{64}, 251-259 (1995).
\bibitem{Schrein-1924} F. A. H. Schreinemakers, G. BerkhoR, and K. Posthumus, \textit{Le Syst$\grave{e}$me: Cu(NO$_3$)$_2$-NH$_4$NO$_3$-H$_2$O}, Recueil des Travaux Chimiques des Pays-Bas \textbf{43}, 508-511 (1924).
\bibitem{Tachiki-1970} M. Tachiki and T. Yamada, \textit{Spin Ordering and Thermodynamical Properties in Spin-Pair Systems under Magnetic Fields}, Prog.Theor.Phys.Suppl. \textbf{24}, 291 (1970).

% [paw]
\bibitem{paw}
P. E. Bl\"{o}chl, Phys. Rev. B \textit{Projected augmented-wave method} $\bf 50$, 17953 (1994); G. Kresse and D. Joubert, Phys. Rev. B \textit{From ultrasoft pseudopotentials to the projector augmented-wave method} $\bf 59$, 1758 (1999).
%[VASP]
\bibitem{Kresse-1993} G. Kresse, et al., \textit{Ab Initio Molecular Dynamics for Open-Shell Transition Metals}, Phys. Rev. B 48, 13115(1993); G. Kresse, et al., \textit{Efficiency of ab-initio total energy calculations for metals and semiconductors using a plane-wave basis set}, Comput. Mater. Sci. 6, 15 (1993).
%[GGA]
\bibitem{Perdew-1996} J.P. Perdew, et al., Generalized Gradient Approximation Made Simple, Phys. Rev. Lett. 77, 3865(1996).

\bibitem{TMRGApp} X. Wang and L. Yu, \textit{Magnetic-Field Effects on Two-Leg Heisenberg Antiferromagnetic Ladders: Thermodynamic Properties}, Phys. Rev. Lett. \textbf{84}, 5399 (2000); J. Lou, T. Xiang, and G. Su, \textit{Thermodynamics of the Bilinear-Biquadratic Spin-One Heisenberg Chain}, Phys. Rev. Lett. \textbf{85}, 2380 (2000); B. Gu and G. Su,  \textit{Comment on ``Experimental Observation of the 1=3 Magnetization Plateau in the Diamond-Chain Compound Cu$_3$ (CO$_3$)$_2$OH$_2$"}, Phys. Rev. Lett. \textbf{97}, 089701 (2006).
\bibitem{Yan-2012} X. Yan, W. Li, Y. Zhao, S.-J. Ran, and G. Su, \textit{Phase diagrams, distinct conformal anomalies, and thermodynamics of spin-1 bond-alternating Heisenberg antiferromagnetic chain in magnetic fields}, Phys. Rev. B \textbf{85}, 134425 (2012).
\bibitem{Ran-2012} S.-J. Ran, W. Li, B. Xi, Z. Zhang, and G. Su, \textit{Optimized decimation of tensor networks with super-orthogonalization for two-dimensional quantum lattice models}, Phys. Rev. B \textbf{86}, 134429 (2012).

\bibitem{FootnoteKappa} During the process of contracting thermal tensor networks, the tensor elements in MPO becomes larger and larger as temperature continues to cool down, and eventually diverges. Therefore, after each single substep of projecting $\nu$ tensors to MPO, one need to extract some renormalization factor $\kappa$ from the MPO to avoid divergence, see more details in Ref. \onlinecite{LTRG}.

\bibitem{Verstraete} F. Verstraete and J.I. Cirac, \textit{Renormalization algorithms for Quantum-Many Body Systems in two and higher dimensions}, cond-mat/0407066 (unpublished).
\bibitem{Feiguin-2005} A.E. Feiguin and S.R. Whitep, \textit{Finite-temperature density matrix renormalization using an enlarged Hilbert space}, Phys. Rev. B \textbf{72}, 220401(R) (2005).

% footnote
\bibitem{Footnote1} The experimental data in Fig. \ref{Fig:SpecificHeat}(d) is taken from Ref. \onlinecite{Bonner-1983} (Fig. 12), where the diverging peak at around 0.18 K revealing a 3D magnetic phase transition is not included due to its irrelavence to the 1D model fitting.
\bibitem{FootnoteCp} The fittings of specific heat curves at $B=2.82$ and 3.57 T in Ref. \onlinecite{Bonner-1983} are based on ED results of spin chain with length up to 12 (and extrapolations therein), which show quite conspicuous discrepancy from the experimental results, and are different from our $\alpha=0.27$ fitting here, especially at low temperatures. We ascibe this difference to the insufficiently small system sizes adopted in Ref. \onlinecite{Bonner-1983} to simulate gapless Luttinger liquid phase at $B = 2.82$ and 3.57 T.
\bibitem{Footnote3} This ideal magnetization curve is calculated from 1D Hamiltonian Eq. (\ref{Eq-Hamiltonian}) where interchain couplings are ignored. This curve is plotted just for elaborating two quantum critical points in the course of applying magnetic fields, and might have some distiction from realistic magnetization curve of CN since the interchain coupling might have some significant influence on the curve at such low temperatures ($<100$ mK).

\end{thebibliography}
\end{document}